\documentclass[twocolumn,amsmath,amssymb,floatfix,prl,footinbib,superscriptaddress]{revtex4-1}

\usepackage{amsmath,amssymb,natbib,bm,graphicx,url,epsf,color}
\usepackage[english]{babel}
\usepackage{wasysym}
\usepackage{MnSymbol}
\usepackage{amsmath}
\usepackage{graphicx}
\usepackage{pdfpages}
\usepackage[colorinlistoftodos]{todonotes}
\usepackage[colorlinks=true, allcolors=blue]{hyperref}
\usepackage[labelfont=bf,justification=raggedright,labelsep = space]{caption}

\newcommand{\QH}{QH~}

\newcommand{\IS}{I_{\mathrm{S}}}
\newcommand{\Is}{I_{\mathrm{S}}}
\newcommand{\ID}{I_{\mathrm{D}}}
\newcommand{\Id}{I_{\mathrm{D}}}
\newcommand{\Vbg}{V_{\mathrm{BG}}}

\newcommand{\Jq}{J_{\mathrm{Q}}}

\newcommand{\Jqe}{J_{\mathrm{Q}}^\mathrm{e}}

\newcommand{\Ts}{T_{\mathrm{S}}}

\newcommand{\Td}{T_{\mathrm{D}}}
\newcommand{\Tzero}{T_{\mathrm{0}}}

\newcommand{\DeltaTs}{\Delta T_{\mathrm{S}}}

\newcommand{\DeltaTd}{\Delta T_{\mathrm{D}}}
\newcommand{\kB}{k_{\mathrm{B}}}

\newcommand{\Gcorb}{G_{\mathrm{corb}}}
\newcommand{\nuc}{\nu_{\mathrm{c}}}

\makeatletter
\AtBeginDocument{\let\LS@rot\@undefined}
\makeatother

\makeatletter
\def\NAT@spacechar{}
\makeatother

\begin{document}
\title{Vanishing bulk heat flow in the $\nu=0$ quantum Hall ferromagnet in monolayer graphene}

\author{R. Delagrange}\thanks{These authors contributed equally to this work.}
\affiliation{Universit\'e Paris-Saclay, CEA, CNRS, SPEC, 91191 Gif-sur-Yvette cedex, France
}
\author{M. Garg}\thanks{These authors contributed equally to this work.}
\affiliation{Universit\'e Paris-Saclay, CEA, CNRS, SPEC, 91191 Gif-sur-Yvette cedex, France
}
\author{G. Le Breton}
\affiliation{Universit\'e Paris-Saclay, CEA, CNRS, SPEC, 91191 Gif-sur-Yvette cedex, France
}
\author{A. Zhang}
\affiliation{Universit\'e Paris-Saclay, CEA, CNRS, SPEC, 91191 Gif-sur-Yvette cedex, France
}
\author{Q. Dong}
\affiliation{CryoHEMT, 91400 Orsay, France
}
\author{Y. Jin}
\affiliation{Universit\'e Paris-Saclay, CNRS, Centre de Nanosciences et de Nanotechnologies (C2N), 91120 Palaiseau, France
}
\author{K. Watanabe}
\affiliation{Research Center for Materials Nanoarchitectonics, National Institute for Materials Science, 1-1 Namiki, Tsukuba 305-0044, Japan
}
\author{T. Taniguchi}
\affiliation{Research Center for Materials Nanoarchitectonics, National Institute for Materials Science, 1-1 Namiki, Tsukuba 305-0044, Japan
}
\author{P. Roulleau}
\affiliation{Universit\'e Paris-Saclay, CEA, CNRS, SPEC, 91191 Gif-sur-Yvette cedex, France
}
\author{O. Maillet}
\affiliation{Universit\'e Paris-Saclay, CEA, CNRS, SPEC, 91191 Gif-sur-Yvette cedex, France
}
\author{P. Roche}
\affiliation{Universit\'e Paris-Saclay, CEA, CNRS, SPEC, 91191 Gif-sur-Yvette cedex, France
}
\author{F.D. Parmentier}
\affiliation{Universit\'e Paris-Saclay, CEA, CNRS, SPEC, 91191 Gif-sur-Yvette cedex, France
}

\date{\today}

\maketitle

\textbf{Under high perpendicular magnetic field and at low temperatures, graphene develops an insulating state at the charge neutrality point. This state, dubbed $\nu=0$, is due to the interplay between electronic interactions and the four-fold spin and valley degeneracies in the flat band formed by the $n=0$ Landau level. Determining the ground state of $\nu=0$, including its spin and valley polarization, has been a theoretical and experimental undertaking for almost two decades. Here, we present experiments probing the bulk thermal transport properties of monolayer graphene at $\nu=0$, which directly probe its ground state and collective excitations. We observe a vanishing bulk thermal transport, in contradiction with the expected ground state, predicted to have a finite thermal conductance even at very low temperature. Our result highlight the need for further investigations on the nature of $\nu=0$.}

The $\nu=0$ quantum Hall (QH) state of graphene stems from the lifting of spin and valley degeneracies in the $n=0$ Landau level (LL) at half filling due to Coulomb interactions~\cite{Kharitonov2012}. The competition between short-range electron-electron interactions, electron-phonon interactions, Zeeman effect due to the applied magnetic field, and sublattice symmetry breaking due to the hexagonal boron nitride (hBN) substrate gives rise to a rich phase diagram for $\nu=0$ that has been extensively studied theoretically~\cite{Nomura2006,Alicea2006,Fertig2006,Jung2009,Goerbig2011,Kharitonov2012,Wu2014,DeNova2017,Atteia2021,Das2022,Hegde2022}, as well as experimentally~\cite{Abanin2007,Checkelsky2008,Amet2013,Young2014,Zibrov2018,Stepanov2018,Veyrat2020,Fu2021,Liu2022,Coissard2022}. It is composed of four ground states with different spin and valley orders, the latter being equivalent to the sublattice order in the $n=0$ LL. These states are represented in Fig.~\ref{fig1}a as a pair of spins distributed on the two-atoms unit cell of graphene's hexagonal lattice~\cite{Kharitonov2012,Young2014,Zibrov2018,Veyrat2020,Coissard2022}. In the ferromagnet (F) and canted antiferromagnet (CAF) phases, each sublattice is occupied by one spin, with a ferromagnetic (F) and antiferromagnetic (CAF) spin texture. In the fully sublattice polarized phase (FSP, also referred to as charge density wave), only one sublattice is occupied by the two spins, aligned antiferromagnetically. Finally, in the Kekule distorsion (KD) phase, also referred to as Kekule bond order~\cite{Coissard2022} or intervalley coherent phase~\cite{Liu2022}, both spins, aligned antiferromagnetically, have the same sublattice polarization, given by a superposition of the two sublattices. Distinguishing these different ground states has long been an experimental challenge, as the CAF, FSP and KD phases are fully insulating, while the F phase realizes a quantum spin Hall (QSH) insulator, with an insulating bulk and counterpropagating helical edge states. The pioneering observation of a continuous transition between an insulating state and a QSH insulator while applying an in-plane magnetic field has led to propose the CAF as the preferred insulating ground state under perpendicular magnetic field~\cite{Young2014}. Subsequent works have shown that the alignement with the hBN subtrate tends to favor the FSP phase~\cite{Zibrov2018}; however, two recent scanning tunneling microscopy experiments~\cite{Liu2022,Coissard2022} suggest that the preferred ground state is the KD phase. Inspired by predictions made for both monolayer~\cite{Takei2016,DeNova2017} and bilayer graphene~\cite{Pientka2017}, we have addressed this problem by probing the thermal transport properties of $\nu=0$. Indeed, each of the candidate ground state hosts chargeless collective excitations linked to its spin and sublattice texture~\cite{Wu2014,Takei2016,Pientka2017,DeNova2017,Stepanov2018,Wei2021}, which can transport heat across the electrically insulating bulk. The CAF and KD phases host gapless collective modes, as they respectively spontaneously break spin and valley symmetries~\cite{Wu2014}; this leads to a finite bulk thermal conduction even at very low temperatures, reflecting the collective modes' dispersion relation~\cite{Pientka2017}. The collective modes of the F and FSP phases are gapped, leading to a zero bulk thermal conduction at low temperatures (below $1~$K in bilayer graphene~\cite{Pientka2017}). Heat transport measurements thus offer a complementary characterization of the phase diagram of $\nu=0$, that allows probing the chargeless collective modes associated to the different ground states.

\section{Samples and experimental principle}

\begin{figure*}[ht]
\centering
\includegraphics[width=0.94\textwidth]{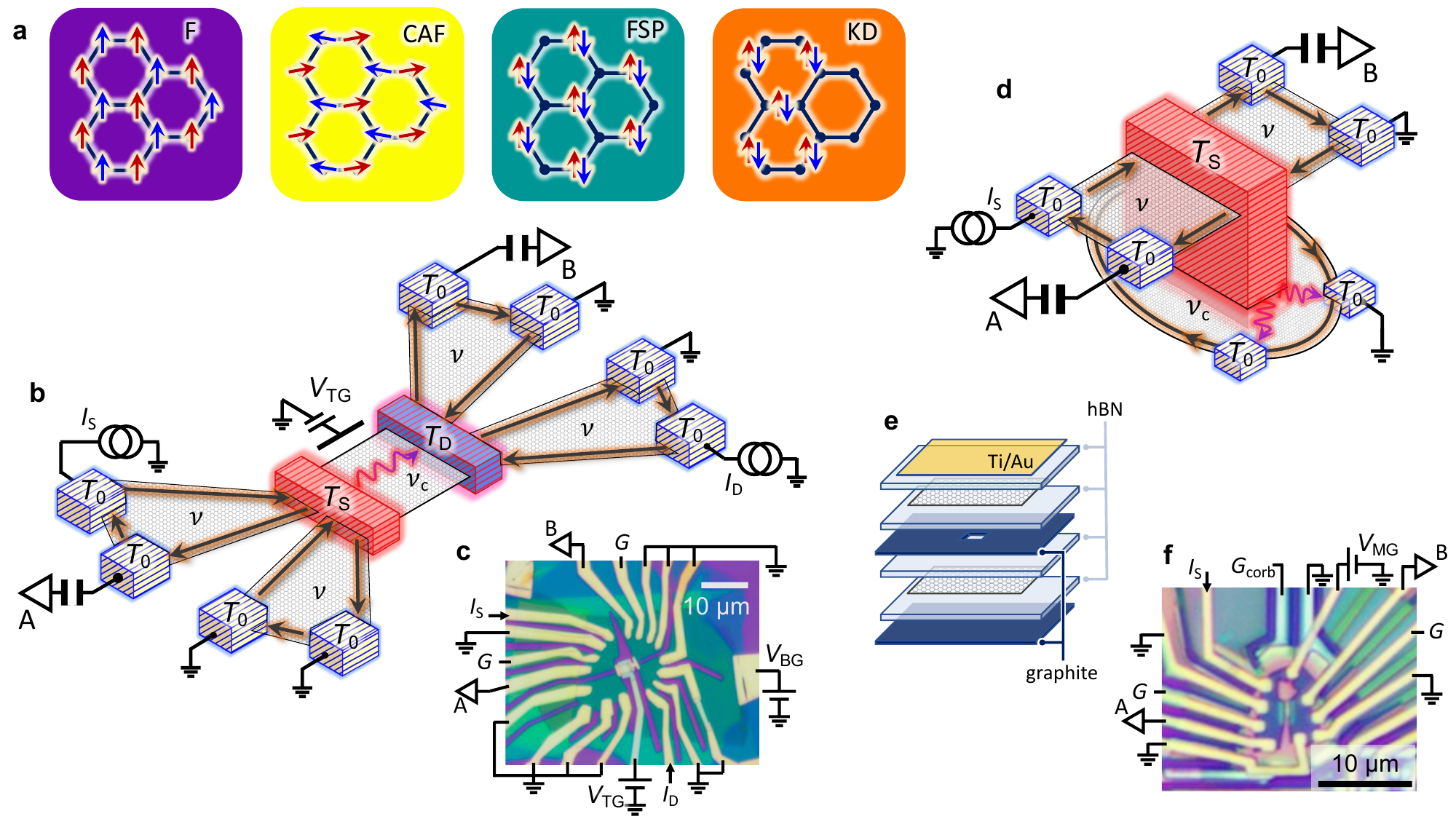}
\caption{\label{fig1} \textbf{$\vert$ $\nu=0$ ground states and graphene heat transport devices. a,} Schematic representation of the spin and valley orders of the four predicted ground states for $\nu=0$. \textbf{b,} Schematic representation of the two-terminal devices. The red and purple bricks represent the source and detector floating metallic contacts with respective temperatures $\Ts$ and $\Td$. The amplifiers A and B represent the two noise measurement lines. \textbf{c,} Optical micrograph of two-terminal device 1, with its wiring configuration. \textbf{d,} Schematic representation of the Corbino heat transport device. The red brick represents the source floating metallic contact with temperature $\Ts$. \textbf{e,} Sketch of the heterostructure realized for the Corbino device, including the graphite back and middle gates, and the metallic top gate. \textbf{f,} Optical micrograph of the Corbino device with its wiring configuration.} 
\end{figure*}

We probed the thermal transport properties of $\nu=0$ using three devices with two specific geometries, presented in Fig.~\ref{fig1}. Both rely on the \QH heat transport technique pioneered in GaAs~\cite{Jezouin2013a}, and recently adapted to graphene~\cite{Srivastav2019,Srivastav2021,Srivastav2022,LeBreton2022}. Micron-size floating metallic contacts are used as hot electrons reservoirs, the temperature of which is controlled by flowing dc electric currents in the sample, and inferred from high sensitivity noise measurements (see methods). The first geometry, depicted in Fig.~\ref{fig1}b, is dubbed \textit{two-terminal} as it uses two floating contacts, source (red brick) and detector (purple brick), with electron temperatures $T_{\mathrm{S}}$ and $T_{\mathrm{D}}$. Similarly to a recent experiment in GaAs~\cite{Melcer2023}, each floating contact is connected to two outer regions of graphene with filling factor $\nu$ that allow independently controlling and measuring its temperature. These regions are connected to larger, cold contacts (blue bricks) at base electron temperature $T_0$, connected to the experiment wiring as shown in Fig.~\ref{fig1}c. The total heat flow carried by the $2\nu$ edge channels of the outer regions is given by $2\nu\frac{\kappa_0}{2}(T_\mathrm{S/D}^2-T_0^2)$~\cite{Srivastav2019,Srivastav2021,Srivastav2022,LeBreton2022}, with $\frac{\kappa_0}{2}(T_\mathrm{S/D}^2-T_0^2)$ the universal quantized heat flow in a one-dimensional ballistic channel~\cite{Pendry1983,Rego1999,Schwab2000,Meschke2006,Jezouin2013a,Banerjee2017} ($\kappa_0=\frac{\pi^2 k_{\mathrm{B}}^2}{3h}$; $h$ is Planck's constant and $\kB$ Boltzmann's constant). The two contacts are connected to a central region through which they can exchange charge and heat. A global back gate and local top gate (see Methods) allow us to independently tune $\nu$ and the filling factor $\nuc$ in the central region. Setting $\nu$ to 1 or 2, and $\nuc=0$, we thus heat up the source while simultaneously measuring the temperature of the detector $T_{\mathrm{D}}$. We also measure the electrical conductance across the device, making sure that no electrical current flows between source and detector. The bulk thermal conductance of $\nuc=0$ is then revealed by a finite increase $\DeltaTd$. Two devices (1 and 2) were fabricated in this geometry in hBN-encapsulated graphene heterostructures. Fig.~\ref{fig1}c shows an optical micrograph of device 1. Importantly, in that device, the graphene flake was found to be aligned with one of the hBN crystals, with an angle mismatch lower than $1^\circ$, giving rise to an insulating state at the charge neutrality point at zero magnetic field~\cite{SM,Zibrov2018}. The experimental wiring is also shown in Fig.~\ref{fig1}c, including the noise measurement lines (A and B), the two current feeds allowing heating either the source or the detector, conductance measurements lines (\textit{G}) used to characterize each region of the sample~\cite{SM}, cold grounds (the cold grounds in between noise measurement lines and current feeds are not shown in Fig.~\ref{fig1}b and d), and the connections of both top and back gates. Note that a small leakage was present on the back gate~\cite{SM} of device 1, restricting the values to $\Vbg\sim 0$).

\begin{figure*}[ht]
\centering
\includegraphics[width=0.99\textwidth]{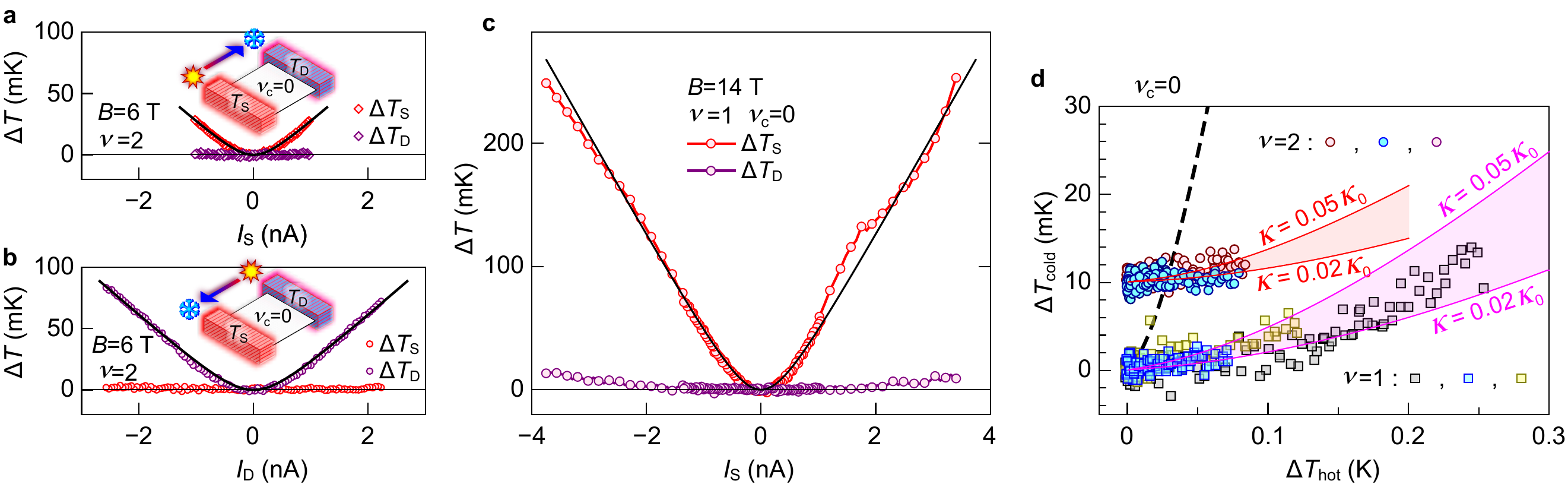}
\caption{\label{fig2} \textbf{$\vert$ Two terminal heat transport at $\nuc=0$. a} and \textbf{b,} Source contact ($\DeltaTs$, red symbols) and detector contact ($\DeltaTd$, purple symbols) temperature increases versus dc current, measured in two-terminal device 1 at 6 T (outer filling factor $\nu=2$, central part filling factor $\nuc=0$). In \textbf{a,} the source is heated up (see schematics in inset) by increasing the source current $\Is$; in \textbf{b,} the detector is heated up by increasing $\Id$. Black line: electronic heat transport model (see text) with $\Tzero\approx 20~$mK. \textbf{c,} $\DeltaTs$ (red) and $\DeltaTd$ (purple) versus $\Is$, measured at 14 T (outer filling factor $\nu=1$, central part filling factor $\nuc=0$). Black line: electronic heat transport model with $\Tzero\approx 45~$mK. \textbf{d,} $\Delta T$ in the colder contact versus $\Delta T$ in the hotter contact. Circles: data at 6 T ($\nu=2$), shifted vertically by $10~$mK fo clarity; squares: data at 14 T ($\nu=1$). Each color corresponds to a different dataset. Full lines: $\Delta T$ yielded by an electron-like heat flow with prefactor $\kappa\approx 0.02-0.05 \kappa_0$; black dashed line: $\Delta T$ yielded by the predicted CAF bulk heat flow in bilayer graphene (see text).} 
\end{figure*}

The second, complementary, geometry is depicted in Fig.~\ref{fig1}d and is dubbed \textit{heat Corbino}. A single floating metallic contact (red brick) with temperature $T_\mathrm{S}$ vertically connects two graphene flakes embedded in a multi-level heterostructure: the top graphene realizes the standard heat transport geometry, with two separate parts allowing to control and measure $T_\mathrm{S}$ (respectively by applying a current bias $\Is$ and measuring the noise through lines A and B, see methods); the bottom graphene has a Corbino-like geometry, with the floating contact connected solely to its bulk, and two electrodes connected to its edge, one of them being connected to the cold ground. The filling factors of the top ($\nu$) and bottom ($\nuc$) graphene flakes are controlled independently using three gates. By comparing the temperature increase $\DeltaTs$ at finite $\Is$ for different values of $\nuc$, we can infer whether there is bulk heat transport in the bottom graphene. In particular, we compare $\nuc=0$ with $\nuc=2$, the latter being expected to have negligible bulk transport~\cite{Melcer2023b}. This approach allows discarding the potential contribution of parallel heat transport channels independent of $\nuc$, such as phonons. Fig.~\ref{fig1}e shows a schematic representation of the heterostructure used to implement this geometry, including the two graphene flakes, the graphite back and middle gate (the latter screening Coulomb interactions between the graphene flakes), and the metallic top gate, each separated by a $\sim40~$nm thick hBN crystal~\cite{SM}. An optical micrograph of the corresponding sample is shown in Fig.~\ref{fig1}f, with the two topmost connections corresponding to the bottom graphene. The connection labeled $\Gcorb$ allows measuring both the two-points edge conductance of the bottom graphene and its bulk conductance through the floating contact.

\section{Two-terminal device}

\begin{figure*}[ht]
\centering
\includegraphics[width=0.98\textwidth]{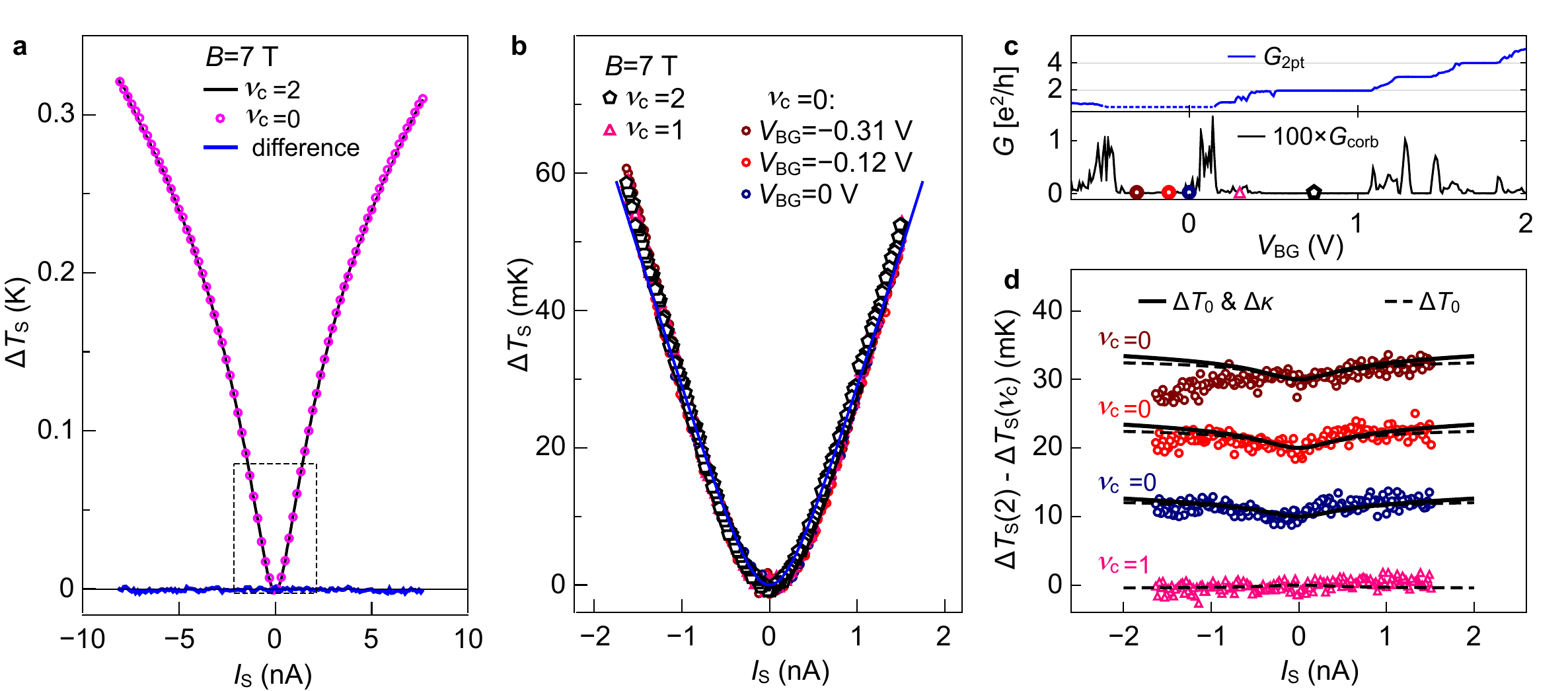}
\caption{\label{fig3} \textbf{$\vert$ $\nuc=0$ heat transport in the Corbino device. a,} Temperature increase $\DeltaTs$ versus $\Is$ measured at 7 T ($\nu=2$) for the filling factor of the lower graphene set to $\nuc=2$ (black line) and to $\nuc=0$ (magenta circles). The blue line is the difference $\DeltaTs(\nuc=2)-\DeltaTs(\nuc=0)$. \textbf{b,} $\DeltaTs$ versus $\Is$ in measured in a smaller span of $\Is$, indicated by the dashed rectangle in \textbf{a}. The symbols correspond to different filling factors of the lower graphene: $\nuc=2$ (white pentagons), $\nuc=1$ (pink triangles), and $\nuc=0$ (circles; the colors correspond to different gate voltages on the $\nuc=0$ plateau, indicated in \textbf{c}). Blue line: electronic heat transport model with $\Tzero\approx 15~$mK. \textbf{c,} Conductances of the lower graphene versus $\Vbg$. Top panel: two-point conductance along the edge (blue). The dashed line corresponds to the $\nuc=0$ plateau where the two-point resistance measurement saturates (see methods). Bottom panel: bulk conductance (black). The symbols indicate the gate voltages corresponding to the data in \textbf{b}. \textbf{d,} Difference $\DeltaTs(\nuc=2)-\DeltaTs(\nuc=0)$ of the data in \textbf{b}. The curves at $\nuc=0$ are incrementally shifted by $10~$mK for clarity. The dashed lines correspond to a difference in $T_0$ smaller than $3~$mK between the data at $\nu=2$ and the other datasets. The full lines correspond to a small difference in $T_0$ combined with an additional heat flow (see text).} 
\end{figure*}

Fig.~\ref{fig2} shows typical measurements obtained with two-terminal device 1. We plot in Fig.~\ref{fig2}a and b measurements of the source and detector temperatures increase as a function of $\IS$ (Fig.~\ref{fig2}a), and $\ID$ (Fig.~\ref{fig2}b). The magnetic field is $B=6~$T, with the filling factor of the outer parts set to $\nu=2$, and that of the central part set to $\nuc=0$. Both measurements show that while we increase the temperature of the source (resp. detector) by $50-100~$mK, the temperature of the detector (resp. source) scarcely changes. In addition, the $\Delta T_{\mathrm{S}/\mathrm{D}}(I_{\mathrm{S}/\mathrm{D}})$ behavior of the heated contact is well matched by a heat transport model invoking only the edge channels of the $\nu=2$ outer regions~\cite{SM}, with a base electron temperature $T_0\sim20~$mK (black line). This suggests that heat transport in the $\nuc=0$ region is much smaller than that of the $\nu=2$ edge at these magnetic field and temperatures. The sensitivity to the $\nuc=0$ heat flow can be increased by setting the outer parts to $\nu=1$, minimizing cooling of the detector contact by edge heat transport. Fig.~\ref{fig2}c shows the results of this measurement in that same device, at $B=14$~T to reach $\nu=1$, on a larger range of $\Is$. While a finite increase $\DeltaTd\approx5-10~$mK is observed above $\Is=2$~nA, the corresponding source temperature is much higher, $\DeltaTs\leq200~$mK, and again well described by an edge heat transport model with $T_0\sim45~$mK (black line), confirming the $B=6~$T result. To estimate the magnitude of the heat flow at $\nuc=0$, we plot in Fig.~\ref{fig2}d the temperature increase of the cold (\textit{e.g.} detector) contact versus that of the heated (\textit{e.g.} source) contact. We observe a similar behavior at $B=6~$T ($\nu=2$) and $B=14~$T ($\nu=1$). Quantitatively, the small temperature increase can be fitted (red and magenta lines) using the simple heat balance:

\begin{equation}
\frac{\kappa}{2}(T_\mathrm{hot}^2-T_\mathrm{cold}^2)=2\nu\frac{\kappa_0}{2}(T_\mathrm{cold}^2-T_0^2),
\label{eq:2termheatbalnu0}
\end{equation}

where the right handside of the balance corresponds to electronic edge heat flow in the outer part of the cold contact, while the left handside models the heat flow in the $\nuc=0$ region with a prefactor $\kappa$ that has the same dimension as $\kappa_0$. The fits yields $\kappa\approx 0.02-0.05 \kappa_0$, confirming the very small magnitude of the bulk heat flow at $\nuc=0$. In contrast, the temperature increase due to a bulk heat flow $J_\mathrm{CAF}\approx6\times 10^{-12}(T_\mathrm{S}^3-T_\mathrm{D}^3)$, predicted for the CAF phase in bilayer graphene~\cite{Pientka2017,SM}, shown as a black dashed line, is much larger than our measurements. This discrepancy is further discussed in the final section of this article. Measurements in device 2, in which the graphene flake was not aligned with the hBN substrate, yielded similar results, shown in the Supplementary Information~\cite{SM}. Note that it is possible that the measured heat flow corresponds to other degrees of freedom, such as phonons. As we explain below, the Corbino heat transport geometry allows circumventing this issue.

\section{Heat Corbino device}

\begin{figure}[ht]
\centering
\includegraphics[width=0.45\textwidth]{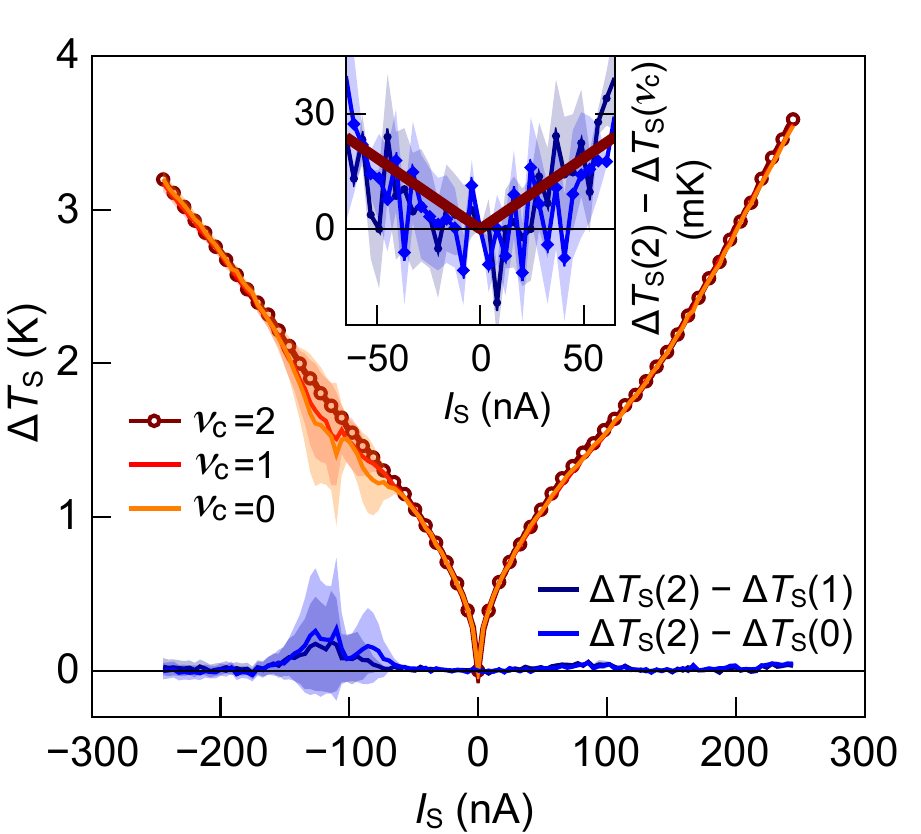}
\caption{\label{fig4} \textbf{$\vert$ Large bias heat transport in the Corbino device.} $\DeltaTs$ versus $\Is$ measured at 12 T ($\nu=1$) for $\nuc=2$ (dark red circles), $\nuc=1$ (red line), and $\nuc=0$ (orange line). Blue line: difference $\DeltaTs(\nuc=2)-\DeltaTs(\nuc=0)$; dark blue line: difference $\DeltaTs(\nuc=2)-\DeltaTs(\nuc=1)$. The shaded areas indicate the error bars on the measurement due to uncertainties in the noise calibration (see methods). Inset: zoom on the temperature differences for $\vert\Is\vert<70~$nA. The dark red line corresponds to an additional thermal flow with $\kappa\sim 0.02 \kappa_0$.}
\end{figure}

Typical measurements in the heat Corbino device are shown in Fig.~\ref{fig3}. Fig.~\ref{fig3}a shows the temperature increase $\DeltaTs$ of the single floating contact versus the dc current $\IS$ fed into the top graphene, the filling factor of which set to $\nu=2$ at 7~T. The data at $\nuc=2$ (magenta circles) and $\nuc=0$ (black line) fall on top of each other, and their difference $\DeltaTs(\nuc=2)-\DeltaTs(\nuc=0)$ (blue line) is zero within our experimental accuracy, up to relatively high current $\IS\approx8~$nA. Fig.~\ref{fig3}b shows high-resolution measurements at low $\IS$ (corresponding to the area indicated as dashed lines in Fig.~\ref{fig3}a) for $\nuc=2$, $1$, and three different back gate voltages $\Vbg$ on the $\nuc=0$ plateau, indicated in the conductance traces of Fig.~\ref{fig3}c (the conductance is expressed in units of the electrical conductance quantum $G_0=e^2/h$, with $e$ the electron charge). Again, the data for all $\nuc$ essentially collapse on a single curve well matched by a heat transport model invoking only the edge channels of the top graphene at $\nu=2$ with $T_0\approx15~$mK (blue line). The difference $\DeltaTs(\nuc=2)-\DeltaTs(\nuc)$ between the $\nuc=2$ data and the other is plotted in Fig.~\ref{fig3}d. A finite positive signal is observed at $\nuc=0$, which could be interpreted as a small ($\kappa\sim 0.1 \kappa_0$) bulk heat flow at $\nuc=0$ (full lines). However, it can also be reproduced by only invoking edge channel heat transport, with a small difference in the base electron temperature $T_0$ of a few mK between the datasets (dashed lines) which were taken over a span of several days.

\begin{figure}[ht]
\centering
\includegraphics[width=0.45\textwidth]{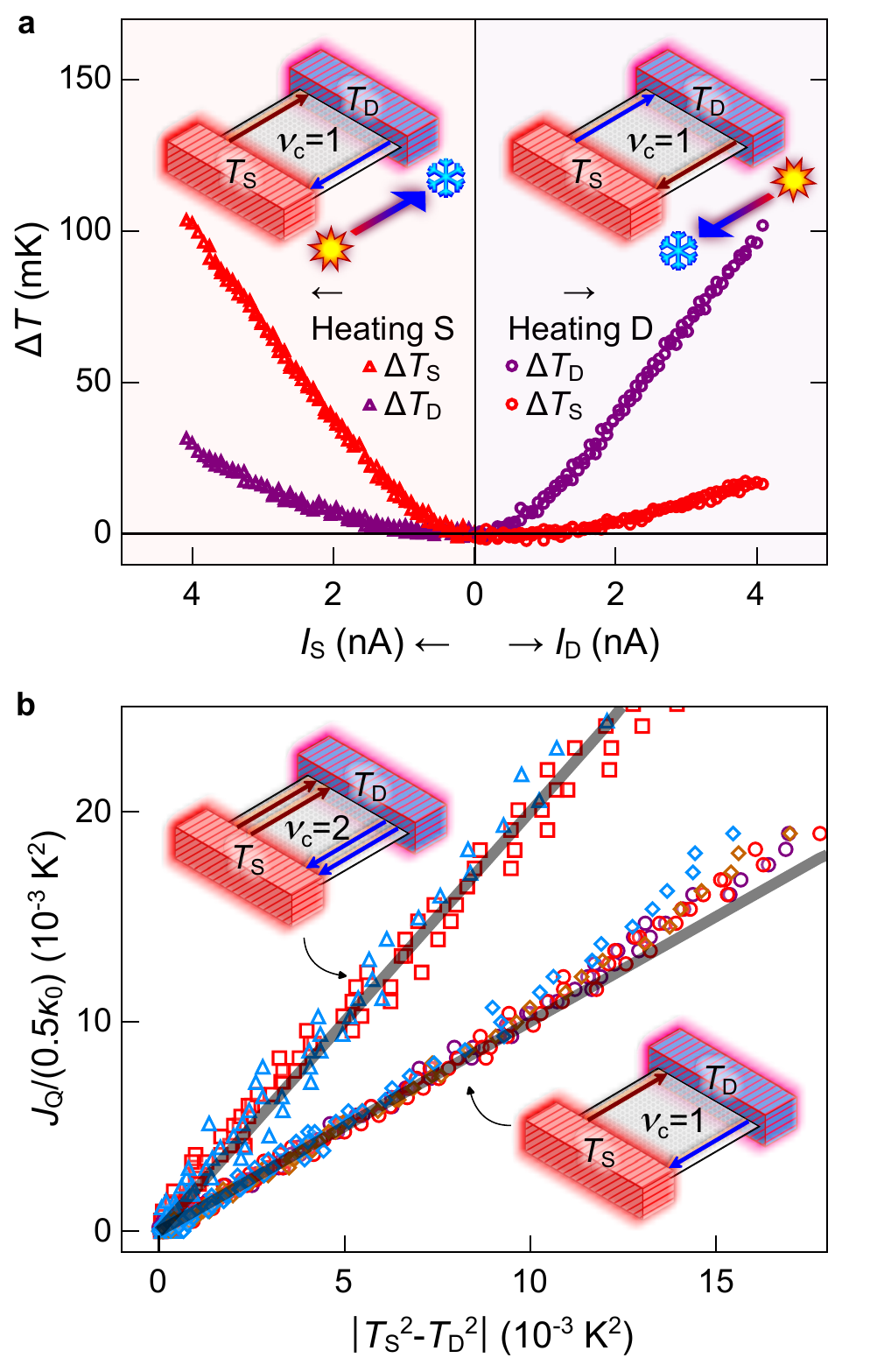}
\caption{\label{fig5} \textbf{$\vert$ Two-terminal quantized edge heat transport in device 1. a,} $\DeltaTs$ (red) and $\DeltaTd$ (purple) versus $\Is$ (left panel, triangles) and $\Id$ (right panel, circles), at 6 T for $\nuc=1$. The insets sketches indicate which contact is primarily heated up. \textbf{b,} Heat flow between the source and detector at 6 T as a function of the difference of their square temperatures $\vert T_\mathrm{S}^2-T_\mathrm{D}^2 \vert$, for $\nuc=1$ (circles and diamonds) and $\nuc=2$ (squares and triangles). Each color and shape corresponds to a specific heating configuration and base $T_0$. Purple (resp. red) circles: bias $\Is$ (resp. $\Id$), $T_0=45-50~$mK. Dark orange (resp. blue) diamonds: bias $\Is$, (resp. $\Id$) $T_0=70-80~$mK. Red squares: bias $\ID$, $T_0=90~$mK. Blue triangles: bias $\ID$, $T_0=110~$mK. Lines: quantized heat flow at $\nuc=1$ and $\nuc=2$, indicated by inset sketches.}
\end{figure}

Because it involves differential temperature measurements of a single floating contact, the heat Corbino geometry is more resolution limited than the two-terminal geometry, such that no conclusive signal can be measured in the few nA range of $\IS$ usually applied in \QH heat transport. Nonetheless, the differential measurements allow subtracting the additional cooling mechanisms independent of $\nuc$: chiefly, the electron-phonon cooling in the floating contact~\cite{Jezouin2013a}, indicated by the sub-linear behavior of $\DeltaTs(\Is)$ in Fig.~\ref{fig3}a. We can thus explore a much larger range of $\IS$, pushing $\DeltaTs$ in the Kelvin range. Fig.~\ref{fig4} shows a measurement of $\DeltaTs(\Is)$ at 12~T. We set $\nu=1$ in the top graphene, where $\DeltaTs$ is larger for a given $\IS$ than at $\nu=2$~\cite{LeBreton2022}. $\DeltaTs(\Is)$ is nearly identical for $\nuc=2$, $1$, and $0$; the sublinear behavior observed above 5~nA in Fig.~\ref{fig3}a persists up to $\vert \IS \vert\approx100~$nA, indicating that cooling is dominated by phonons in the floating contact. We recover a linear behavior at $\Is>100~$nA; note that there, the potential developing in the top graphene is comparable to the Zeeman energy, leading to the emission and absorption of magnons in the top graphene at $\nu=1$~\cite{Wei2018,SM}. This affects both conductance and noise measurements~\cite{SM}, see \textit{e.g.} the increased error bars around $-100~$nA, making our analysis in terms of pure heat transport invalid. As in Fig.~\ref{fig3}, the difference $\DeltaTs(\nuc=2)-\DeltaTs(\nuc)$ (blue and dark blue lines) is essentially zero; nonetheless, a zoom in the phonon-dominated regime before $\nu=1$ magnons are emitted (inset of Fig.~\ref{fig4}) reveals a finite signal of a few tens of mK. Contrary to Fig.~\ref{fig3}d, this signal cannot be fully accounted for by a difference in $T_0$~\cite{SM}, but is comparable to the difference in $\DeltaTs$ caused by an additional bulk thermal flow $\frac{\kappa}{2}(T_\mathrm{S}^2-T_0^2)$, with $\kappa\sim 0.02 \kappa_0$ (dark red line), again much smaller than that of the $\nu=2$ edge.

\section{Quantized edge heat transport in two-terminal}

To check the validity of our heat transport approach, we performed measurements in the two-terminal device 1 with the filling factor of the central region set to $\nuc\neq0$. Fig.~\ref{fig5}a shows the result of such a measurement at 6~T with $\nuc=1$, where the temperature increases $\DeltaTs$ and $\DeltaTd$ are measured simultaneously as a function of $\Is$ (left) and $\Id$ (right). On the left panel of Fig.~\ref{fig5}a, the source is primarily heated up by the current $\Is$, as in Fig.~\ref{fig2}a; the temperature of the detector, however, now increases more clearly, with $\DeltaTd \sim 0.2\DeltaTs$. Heating up the detector by increasing $\Id$ instead of $\Is$ yields the opposite result (right panel), verifying the symmetric nature of the device. Combined with the electrical conductance measurements~\cite{SM}, this gives us a full electrical and thermal characterization of the device, such that we can replot the data in Fig.~\ref{fig5}a in terms of the heat balance between source and detector~\cite{SM}:

\begin{equation}
J_\mathrm{Q}=\nuc\frac{G_0}{2}( V_\mathrm{S}^2-V_\mathrm{D}^2)=\nuc\frac{\kappa_0}{2}( T_\mathrm{S}^2-T_\mathrm{D}^2),
\label{eq:2termheatbalnu12}
\end{equation}

where $V_\mathrm{S}$ and $V_\mathrm{D}$ are the electrical potentials of the source and the detector, respectively. 
The result is shown in Fig.~\ref{fig5}b: all data corresponding to a given $\nuc$ (see inset sketches) fall onto a straight line with slope given by $\nuc$ (grey lines), corresponding to the quantized heat flow carried by $\nuc$ ballistic channels. Note that a finite current offset was present in all datasets~\cite{SM}, leading to an increase in the base electron temperature $T_0$ up to $110~$mK.

\section{Magnitude of the heat flow}

Our results suggest an extremely small bulk heat flow at $\nuc=0$, barely within our experimental sensitivity. Both geometries yield similar magnitudes, twenty to fifty times smaller than the universal quantized heat flow at comparable temperatures. Importantly, the simultaneous conductance measurements yield an upper limit for the electrical conductance across $\nuc=0$ of a few $10^{-3}G_0$~\cite{SM}, unlikely to be the source of our observed residual bulk heat flow. The heat flow $J_\mathrm{CAF}$ calculated in the CAF phase for a $5~\mu$m-wide bilayer graphene sample~\cite{Pientka2017} has a similar magnitude at $T_0=25~$mK as the universal quantized heat flow~\cite{SM}. As exemplified in Fig~\ref{fig2}d, it should thus be much larger than our observations. There are obviously caveats to this comparison: we have studied monolayer graphene, in which the preferred ground state for $\nu=0$ is now thought to be a KD~\cite{Liu2022,Coissard2022}. However, it remains a good point of reference, as $\nu=0$ stems from the same physics in monolayer and bilayer graphene, with similar values of the anisotropy energies underpinning its phase diagram~\cite{Kharitonov2012,Kharitonov2012a,DeNova2017}. In both monolayer and bilayer graphene, the CAF phase has gapless collective excitations, as well as the KD phase and its bilayer counterpart, the partially layer polarized (PLP) phase~\cite{Wu2014,Pientka2017,DeNova2017}, the latter being predicted to have a slightly larger thermal conductance at low temperature~\cite{Pientka2017}. We can thus reasonably assume that the expected thermal conductances for the various ground states of $\nu=0$ in monolayer and bilayer graphene are similar. The large discrepancy with our results raises the question whether there is heat conduction due to gapless collective modes at all at $\nuc=0$. This is compounded by our observation of a similar bulk heat flow at $\nuc=1$, where the collective excitations are spin waves with a large gap of $\sim1~\mathrm{K}/$T given by the Zeeman energy~\cite{Atteia2021,Wei2021}, to which our measurements should thus be insensitive. Our results suggest a FSP phase, which is the only one expected to be both an electrical and thermal insulator. While this is consistent in 2-terminal device 1, where the alignment with the hBN substrate should lead to the FSP phase~\cite{Zibrov2018,Liu2022,SM}, it is less so for the other non-aligned devices. Several hypotheses can be considered in order to explain the absence of heat flow. \textit{i)}, The $\nuc=0$ ground state is indeed FSP in our devices, and the difference with the recent scanning tunneling microscopy observations of a KD phase~\cite{Liu2022,Coissard2022} stems from the full hBN encapsulation. \textit{ii)}, The ground state is KD in the non-aligned devices, with a strongly reduced thermal conductance due to the coexistence of domains with different bond orders~\cite{Coissard2022}. \textit{iii)}, The ground state is more complex version of the KD phase. For instance, the inclusion of $U(1)$-symmetry breaking terms is expected to open gaps in the KD phase spectrum~\cite{Peterson2014,Khanna2023}, although their magnitude remains debated. Alternatively, the KD phase has been recently predicted to coexist with other phases~\cite{Das2022}, the collective excitations spectrum of which might be gapped. \textit{iv)}, The observed absence of heat flow is due to extrinsic factors, such as a collective mode quantization due to the finite size of the device~\cite{Fu2021,SM}, or the presence of an interface thermal resistance between the hot electron bath and the collective modes.
Note that an experiment similar to ours~\cite{Kumarnu=0}, concurrently performed in bilayer graphene, also showed a negligible bulk heat flow; the analogy between monolayer and bilayer graphene then makes hypothesis \textit{i)} less likely. Indeed, the bilayer graphene equivalent of the FSP phase (fully layer polarized, FLP) can be obtained by applying an electric displacement field, which was shown to have no effect on thermal transport~\cite{Kumarnu=0}. Hypothesis \textit{ii)} would intuitively lead to a reduction of the thermal transport due to multiple scatterings on the domains, without necessarily suppressing it. Hypothesis iv) is also less likely: the collective modes quantization was only observed and bilayer graphene~\cite{Fu2021,Wei2018,SM}, and might be suppressed by the bond order domains of \textit{ii)} or the coexistence with other phases of \textit{iii)}. Since the collective modes are electron-hole type excitations~\cite{Pientka2017,Assouline2021,SM}, they are likely to
be well coupled to the hot electron bath, such that the interface resistance is probably small. At this stage, hypothesis \textit{iii)} thus seems the most likely; nevertheless, our results show that additional work, both theoretical and experimental, is needed in order to fully understand the nature of $\nu=0$.

\section{acknowledgments}

 This work was funded by the ERC (ERC-2018-STG \textit{QUAHQ}), by the “Investissements d’Avenir” LabEx PALM (ANR-10-LABX-0039-PALM), and by the Region Ile de France through the DIM QUANTIP. K.W. and T.T. acknowledge support from the JSPS KAKENHI (Grant Numbers 21H05233 and 23H02052) and World Premier International Research Center Initiative (WPI), MEXT, Japan. The authors warmly thank A. Das, R. Kumar, S. K. Srivastav, M. Goerbig, T. Jolicoeur, C. Altimiras, R. Ribeiro-Palau, Y. Hong and P. Jacques for enlightening discussions as well as technical advice and support. 

\section{Data and code availability}

The data shown here and the analysis codes are available on Zenodo: \href{https://doi.org/10.5281/zenodo.10528559}{https://doi.org/10.5281/zenodo.10528559}

\section{Methods}

\subsection{Devices fabrication}

The devices are based on hBN encapsulated graphene stacks (typical hBN thickness: $25-40~$nm), including a graphite back gate. Side metallic contacts are made by $\mathrm{CHF}_3 / \mathrm{0}_2$ reactive ion etching of the stacks down to approximately half the bottom BN, followed by Cr/Au metal deposition in a few $10^{-8}~$mbar vacuum. The stacks are then further etched to define the device geometry. The top gates are realized by transferring a hBN flake on top of the device which we use as dielectrics, on top of which we depose the metal (Ti/Au) gate. The Corbino device also includes a graphite middle gate in between the two graphene layers; more details of the fabrication of this device are given in the Supplementary Information~\cite{SM}.

 \subsection{Electrical conductance measurements}

 Measurements were performed in a cryogen-free dilution refrigerator with base temperature 9~mK, under high magnetic fields (up to 14~T) obtained with a superconducting magnet. The electrical conductance was extracted from lock-in measurements at frequencies below 20~Hz. The conductance measurement lines (indicated $G$ and $G_\mathrm{corb}$ in Fig.~\ref{fig1}), as well as the dc current feed lines (indicated $\IS$ and $\Id$ in Fig.~\ref{fig1}) are heavily filtered to obtain low base electron temperatures~\cite{SM}. All transconductances between the various terminals of the devices were measured simultaneously using different frequencies. The two points conductances, primarily used to characterize the quantum Hall state in the different parts of the devices, were obtained by measuring the ac voltage drop on the sample when subjected to an ac current bias. In the insulating $\nu=0$ state, this can lead to large ac voltage drops, thereby saturating the lock-ins (see \textit{e.g.} Fig.~\ref{fig3}c).

\subsection{Thermal transport and noise measurements}

The thermal transport measurements are based on the heat balance on a floating metallic contact, connected to $N$ outgoing ballistic heat channels, whose electron temperature $\Ts$ is raised with respect to the base electron temperature $T_0$ by locally dissipating Joule power $\Jq$ through the application of a dc current $\Is$~\cite{Jezouin2013a,Srivastav2019}. In the steady state, the Joule power is equal to the sum of the quantized outgoing heat flow carried by the ballistic edge channels $N\kappa_0/2(\Ts^2-T_0^2)$, the heat flow due to electron-phonon cooling in the contact~\cite{Jezouin2013a,Srivastav2019}, and the expected heat flow across the bulk of the $\nuc=0$ region. $\Jq$ is calculated from the applied dc current and the Hall conductance of the device (see below), and the temperature increase $\DeltaTs$ is inferred from the thermal current noise generated by the contact $\Delta S$. This thermal noise is detected through two independent measurement lines (A and B in Fig.~\ref{fig1}). We measure both autocorrelations ($\mathrm{A}\times\mathrm{A}$, $\mathrm{B}\times\mathrm{B}$) and crosscorrelation ($\mathrm{A}\times\mathrm{B}$)~\cite{SM}; in the two-terminal sample, the latter is crucial to independently determine $\DeltaTs$ and $\DeltaTd$ when $\nuc\neq0$, see Fig.~\ref{fig5} as well as the Supplementary Information~\cite{SM}. In the Corbino sample, we use crosscorrelation to remove spurious noise contributions stemming from the injection contact~\cite{SM,LeBreton2022} by calculating $\Delta S=((\mathrm{A}\times\mathrm{A}+\mathrm{B}\times\mathrm{B})/2-\mathrm{A}\times\mathrm{B})/2$. The two measurement lines are calibrated on a regular basis (typically weekly), with fluctuations no larger than $2\%$. The typical uncertainty, given by the size of the symbols in the $\DeltaTs$ plots, is given by the averaging time of about 2 minutes per point. The averaging procedure is done by performing typically 15 back and forth $\Is$ sweeps with a 7~s averaging time per point. The sweeps are then averaged, allowing to compensate for slow drifts in the measurements. The large error bars at negative $\Is$ appearing in Fig.\ref{fig4} correspond to the fluctuations in the calibrated gains, and are due to a very large, positively correlated noise, the origin of which is unclear, and beyond the scope of this letter.

The precise expressions for $\Jq$ and $\DeltaTs$ depend on the configuration of the device. Assuming no electrical transport at $\nuc=0$ (which we check using conductance measurement), this case simplifies in both devices to the expressions~\cite{LeBreton2022} $\Jq=\Is^2/(4\nu G_0)$ and $\DeltaTs=\Delta S / (\nu G_0 \kB)$. The black lines in Figs.~\ref{fig2}a, b, c and the blue line in Fig.~\ref{fig3}b correspond to these $\DeltaTs(\Is)$ calculations, neglecting electron-phonon cooling and bulk heat transport through $\nuc=0$. In these calculations the base electron temperature $T_0$ is used as a fit parameter, whose value is compared to that extracted from the noise in absence of dc current. Importantly, $T_0$ essentially affects the low current rounding of the $\DeltaTs(\Is)$, its slope being related to $N$ and $\nu$~\cite{LeBreton2022}.

The black dashed line in Fig.~\ref{fig2}d, corresponding to the predicted detector temperature increase $\DeltaTd$ as a function of the source temperature increase $\DeltaTs$ in presence of a bulk heat flow $J_\mathrm{CAF}$ at $\nuc=0$, was calculated by numerically solving the heat balance $J_\mathrm{CAF}(\Ts,\Td)=2\nu\Jqe(\Td,T_0)$:

\begin{equation}
6\times 10^{-12}(T_\mathrm{S}^3-T_\mathrm{D}^3)=2\nu\frac{\kappa_0}{2}(T_\mathrm{D}^2-T_0^2).
\label{eq:2termheatbalnu0CAF}
\end{equation}

\end{document}


\title{Supplementary Information for "Vanishing bulk heat flow in the $\nu=0$ quantum Hall ferromagnet in monolayer graphene"}

\author{R. Delagrange}
\affiliation{Universit\'e Paris-Saclay, CEA, CNRS, SPEC, 91191 Gif-sur-Yvette cedex, France
}
\author{M. Garg}
\affiliation{Universit\'e Paris-Saclay, CEA, CNRS, SPEC, 91191 Gif-sur-Yvette cedex, France
}
\author{G. Le Breton}
\affiliation{Universit\'e Paris-Saclay, CEA, CNRS, SPEC, 91191 Gif-sur-Yvette cedex, France
}
\author{A. Zhang}
\affiliation{Universit\'e Paris-Saclay, CEA, CNRS, SPEC, 91191 Gif-sur-Yvette cedex, France
}
\author{Q. Dong}
\affiliation{CryoHEMT, 91400 Orsay, France
}
\author{Y. Jin}
\affiliation{Universit\'e Paris-Saclay, CNRS, Centre de Nanosciences et de Nanotechnologies (C2N), 91120 Palaiseau, France
}
\author{K. Watanabe}
\affiliation{Research Center for Materials Nanoarchitectonics, National Institute for Materials Science, 1-1 Namiki, Tsukuba 305-0044, Japan
}
\author{T. Taniguchi}
\affiliation{Research Center for Materials Nanoarchitectonics, National Institute for Materials Science, 1-1 Namiki, Tsukuba 305-0044, Japan
}
\author{P. Roulleau}
\affiliation{Universit\'e Paris-Saclay, CEA, CNRS, SPEC, 91191 Gif-sur-Yvette cedex, France
}
\author{O. Maillet}
\affiliation{Universit\'e Paris-Saclay, CEA, CNRS, SPEC, 91191 Gif-sur-Yvette cedex, France
}
\author{P. Roche}
\affiliation{Universit\'e Paris-Saclay, CEA, CNRS, SPEC, 91191 Gif-sur-Yvette cedex, France
}
\author{F.D. Parmentier}
\affiliation{Universit\'e Paris-Saclay, CEA, CNRS, SPEC, 91191 Gif-sur-Yvette cedex, France
}

\date{\today}


{
\let\clearpage\relax
\maketitle
}

\section{Charge and heat transport calculations}

\subsection{Basic electronic heat balance}

The heat balance in absence of bulk heat flow at $\nuc=0$ (for either geometry) is obtained by first writing the heat flow through a single integer quantum Hall edge channel stemming from an electron reservoir labelled $\alpha$ with chemical potential $\mu+eV_\alpha$ ($\mu$ is the global chemical potential of the sample in absence of dc bias) and temperature $T_\alpha$:

\begin{equation}
        J_\mathrm{out}^\alpha=\frac{1}{h}\int d\epsilon (\epsilon-\mu) \left[ f_\alpha (\epsilon) -\theta(\mu- \epsilon) \right]=\frac{\pi^2\kB^2T_\alpha^2}{6h}+\frac{1}{h}\frac{(eV_\alpha)^2}{2}=\frac{\kappa_0}{2}T_\alpha^2+\frac{G_0}{2}V_\alpha^2,
    \label{eq:heatflow}
\end{equation}

where $f_\alpha (\epsilon)$ is the Fermi function in the reservoir $\alpha$, and $\theta(\epsilon)$ is Heaviside's step function. This formula can be used to obtain the heat balance at integer filling factor $\nu$ in the central metallic island: $2\sum_1^\nu J_\mathrm{out}^c=\sum_1^\nu J_\mathrm{out}^I+\sum_1^\nu J_\mathrm{out}^G$, where $J_\mathrm{out}^I$ (resp. $J_\mathrm{out}^G$) is the heat flow carried by a single edge channel leaving the current feed contact upstream of the metallic island (resp. the upstream grounded contact on the other side of the metallic island), with temperature $T_0$ and chemical potential $\mu+\IS/(\nu G_0)$ (resp. $\mu$). Recalling that $V_S=\IS/(2 \nu G_0)$, this yields:

\begin{equation}
       2\sum_1^\nu\frac{\kappa_0}{2}T_S^2+2\sum_1^\nu \frac{G_0}{2} \left(\frac{\IS}{2\nu G_0}\right)^2 =2\sum_1^\nu\frac{\kappa_0}{2}T_0^2+\sum_1^\nu \frac{G_0}{2} \left(\frac{\IS}{\nu G_0}\right)^2.
    \label{eq:heatflow2}
\end{equation}

Grouping the temperature and dc current-dependent terms on either side of the equation gives:

\begin{equation}
        \sum_1^\nu \frac{\IS^2}{4\nu^2 G_0}  =2\sum_1^\nu\frac{\kappa_0}{2}(T_S^2-T_0^2),
    \label{eq:heatflow3}
\end{equation}
which yields the equation used throughout the main text to compare our data with ballistic edge heat flow in absence of additional cooling processes:

\begin{equation}
\frac{1}{4\nu G_0}\IS^2 = 2 \nu \frac{\kappa_0}{2}(\Tc^2-T_0^2).
\label{eq:heatbalance}
\end{equation}

\subsection{Heat balance in the 2-terminal geometry}

The transconductances and the dc voltages developing on the source and the detector in the 2 terminal geometry are obtained from the current balances on the two floating contacts. Assuming $\Id=0$,

\begin{align}
V_\mathrm{S}(2\nu+\nuc) = \IS/G_0+\nuc V_\mathrm{D},\nonumber\\
V_\mathrm{D}(2\nu+\nuc) = \nuc V_\mathrm{S}.
\label{eq:currbalance2term}
\end{align}

This yields the voltages:
\begin{align}
V_\mathrm{S} = \frac{1}{4\nu G_0}\IS(1+\frac{\nu}{\nu+\nuc}),\nonumber\\
V_\mathrm{D} = \frac{1}{4\nu G_0}\IS(1-\frac{\nu}{\nu+\nuc}).
\label{eq:voltages2term}
\end{align}

The reflected and transmitted differential transconductances $G_{R/T}=G_0\times dI_\mathrm{A/B}/dI_\mathrm{S}$ are then given by:

\begin{align}
G_{R} = \frac{G_0}{4\nu }\IS(1+\frac{\nu}{\nu+\nuc}),\nonumber\\
G_{T} = \frac{G_0}{4\nu }\IS(1-\frac{\nu}{\nu+\nuc}).
\label{eq:transconductances2term}
\end{align}

The reverse transconductances can be obtained by assuming $\Is=0$ at finite $\Id$.

For both $\Id$ and $\Is$ finite, the Joule power flowing between source and detector can be obtained from Eq.~\ref{eq:currbalance2term}:

\begin{align}
I_\mathrm{S}- I_\mathrm{D}= 2G_0(\nu+\nuc)(V_\mathrm{S}-V_\mathrm{D}),\nonumber
\\
I_\mathrm{S}+ I_\mathrm{D}= 2G_0 \nu(V_\mathrm{S}+V_\mathrm{D}),
\label{eq:currsumdiff}
\end{align}

which yields:

\begin{equation}
I_\mathrm{S}^2- I_\mathrm{D}^2= 4G_0^2\nu(\nu+\nuc)(V_\mathrm{S}^2-V_\mathrm{D}^2).
\label{eq:deltaI2todeltaV2}
\end{equation}

Using Eq.~\ref{eq:heatflow}, one can compute the heat flowing between source and drain $\frac{G_0\nuc}{2}(V_\mathrm{S}^2-V_\mathrm{D}^2)$, and express it as a function of $\Is$ and $\Id$ to obtain the plots shown in main text Fig.~5b:

\begin{equation}
J_\mathrm{Q}^{\mathrm{S}\longrightarrow\mathrm{D}}= \frac{\nuc}{8G_0\nu(\nu+\nuc)}(I_\mathrm{S}^2-I_\mathrm{D}^2).
\label{eq:Pjoule2term}
\end{equation}

\section{Fabrication of the heat Corbino device}

The heat Corbino device was made in several successive stacking and fabrication steps. A first stack of graphite / hBN / graphene / hBN / graphite was prepared and deposited onto a silicon oxide substrate with pre-patterned leads and bonding pads. This stack constitutes the lower part of the whole device; its top graphite (which would end up as the middle graphite gate in between the two graphene layers) was patterned using electron beam lithography and oxygen plasma etching, after which the electrical 1D contacts of the bottom graphite where patterned and deposited using the standard 1D contact technique described in the methods section of the main text. After testing these contacts, a second hBN/graphene/hBN stack was deposited onto the device. Two etching and metallization steps were then realized: in the first step, the central contact and the connection to the middle gate were patterned and obtained by $\mathrm{CHF}_3 / \mathrm{0}_2$ RIE etching through the whole top stack (exposing the middle graphite and the central contact already deposited in the bottom stack) then depositing Cr/Au electrodes. In the second step, the outer contacts of the top graphene were made by partially etching through the top stack (so that the edges of the top graphene are exposed, but not the middle graphite) then depositing Cr/Au electrodes. After testing all contacts, including the connection between both graphenes, and checking that there was no leakage between either graphene and the graphite gates, the top graphene was etched to realize the heat transport geometry depicted in the main text. Lastly, the top gate was made by depositing a final hBN flake on top of the whole device using a propylene carbonate film, then depositing a Ti/Au top gate (which does not appear in the micrograph shown in the main text) matching the shape of the top graphene.

\section{Conductance measurements}

\subsection{Setup}

\begin{figure*}[h!]
\centering
\includegraphics[width=0.75\textwidth]{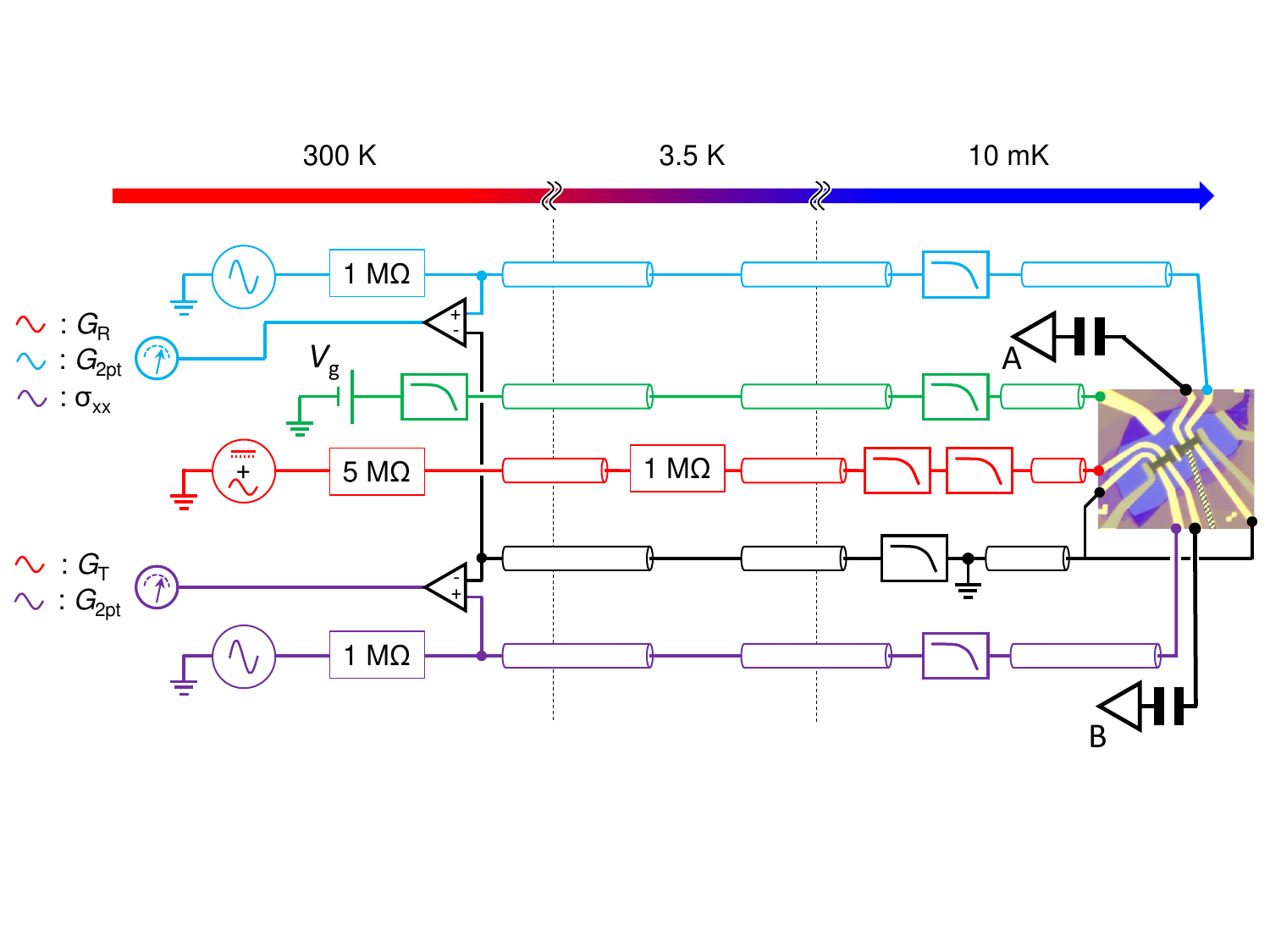}
\caption{\label{figsup-conductances} Layout of the wiring for the conductance measurements. Lines are color-coded (blue: T-side conductance; green: back gate; red: dc current feed; purple: R-side conductance; black: cold ground).}
\end{figure*}

A detailed description of the conductance measurements is shown in Fig.~\ref{figsup-conductances}. The conductances are measured through additional contacts located immediately downstream of the noise measurements contact depicted by the A and B amplifiers in main text Figure 1. The measurements were performed using lock-in techniques at low frequency, below 10~Hz. All lines, including current feed (red in Fig.~\ref{figsup-conductances}) and back gate (green in Fig.~\ref{figsup-conductances}) are heavily filtered at the mixing chamber stage of our dilution refigerator using cascaded \textit{RC} filters. The effect of those filters (both in terms of series resistance and capacitive cutoff) are taken into account in our data. All measurements are performed using differential amplifiers (CELIANS EPC-1B) referenced to the cold ground (black in Fig.~\ref{figsup-conductances}) The latter is directly connected (both electrically and thermally) to the mixing chamber stage. The current feed line includes a $1~$M$\Omega$ series bias resistor thermally anchored to the $3.5~$K stage of our dilution refrigerator. The setup is depicted for a simple thermal transport device (\textit{e.g.} the top part of the heat Corbino device). The transmitted and reflected conductances ($\Gtr$ and $\Gref$ in Fig.~\ref{figsup-conductances}) are defined with respect to the relative positions of the dc current feed contact, the floating metallic island, and the conductance measurement lead. Thus, in the Corbino device, $\Gtr=G_0 dI_B/dI_S$ corresponds to the transconductance of the edge channels flowing from the current feed lead to the floating metallic island, then from the floating contact to the lead connected to amplifier B, while $\Gref=G_0 dI_A/dI_S$ corresponds to the transconductance of the edge channels flowing from the current feed lead to the floating metallic island, then from the floating contact to the lead connected to amplifier A. In the 2-terminal heat transport device, the transmitted conductance refers either to the transconductance $G_0^2 dV_B/dI_S$ of the edge channels flowing from the source current feed lead (where $\IS$ is applied) to the source island, then from the source island to the detector island through the $\nuc=0$ region, then from the detector island to the lead connected to amplifier B, or to the transconductance $G_0^2 dV_A/dI_D$ of the edge channels flowing from the detector current feed lead (where $\Id$ is applied) to the detector island, then from the detector island to the source island through the $\nuc=0$ region, then from the source island to the lead connected to amplifier A.

\subsection{2-terminal device}

\begin{figure}[ht]
\centering
\includegraphics[width=0.4\textwidth]{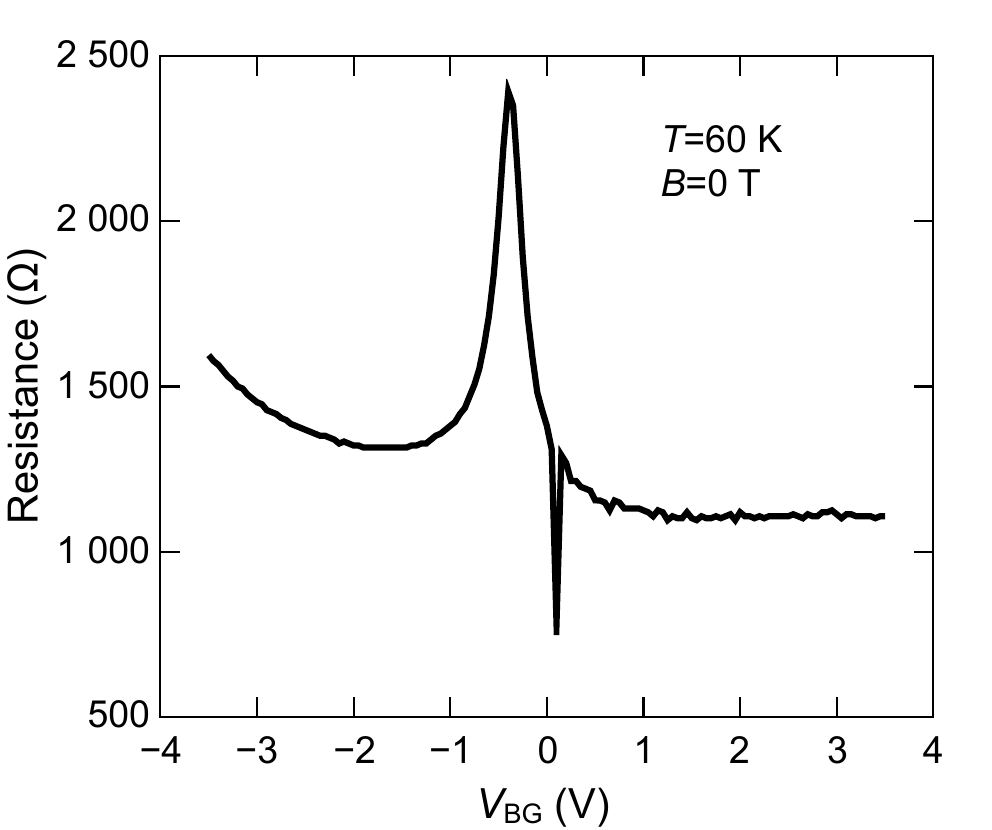}
\caption{\label{figsup-2term_G_zerofield} Measurement of the 2-point resistance versus back gate voltage of the 2-terminal device at low temperature and zero magnetic field. The dip close to zero back gate voltage is a measurement artifact linked to the leaking back gate.}
\end{figure}

\begin{figure}[ht]
\centering
\includegraphics[width=0.8\textwidth]{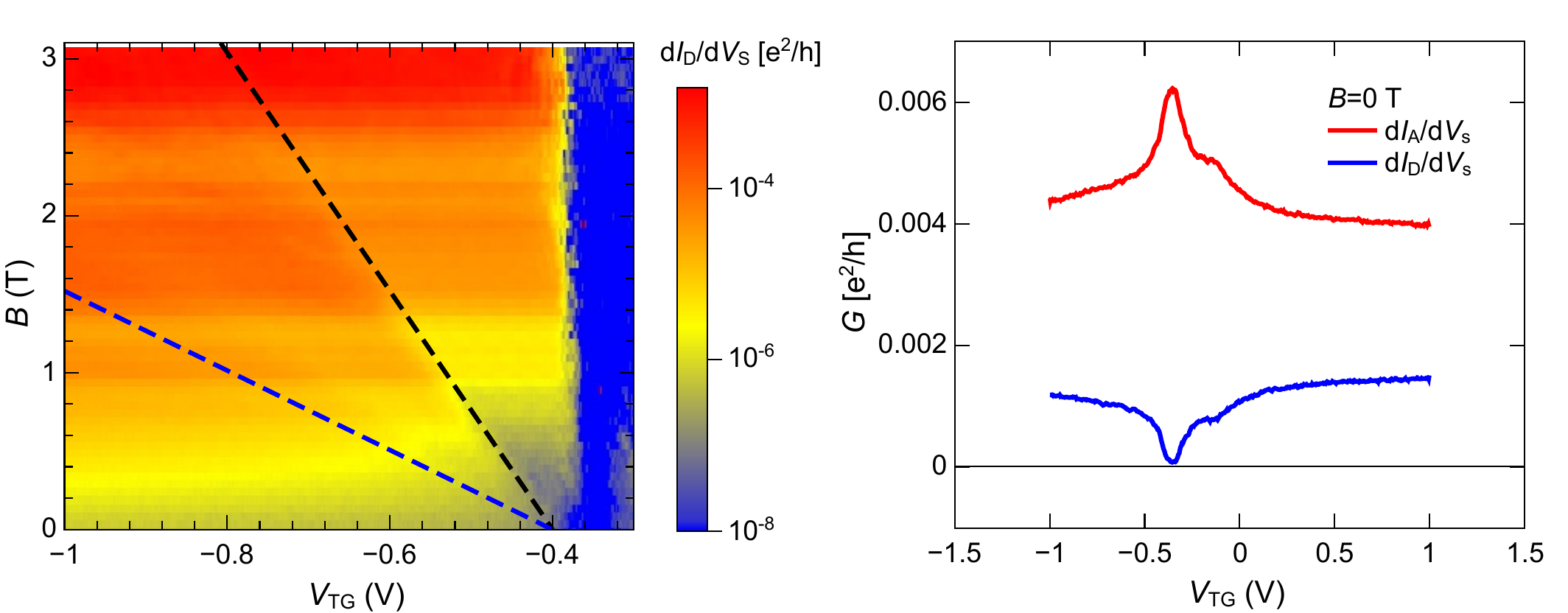}
\caption{\label{figsup-2term_Fandiag} Left: measurement of the electrical conductance between source and detector of the 2-terminal device versus magnetic field and top gate voltage. The \FDP{black} and blue dashed lines correspond to filling factor $\nuc=2$ and $\nuc=6$. Right: conductance between source and detector (blue) and reflected transconductance (red) versus top gate voltage at zero magnetic field.}
\end{figure}

\begin{figure}[ht]
\centering
\includegraphics[width=0.8\textwidth]{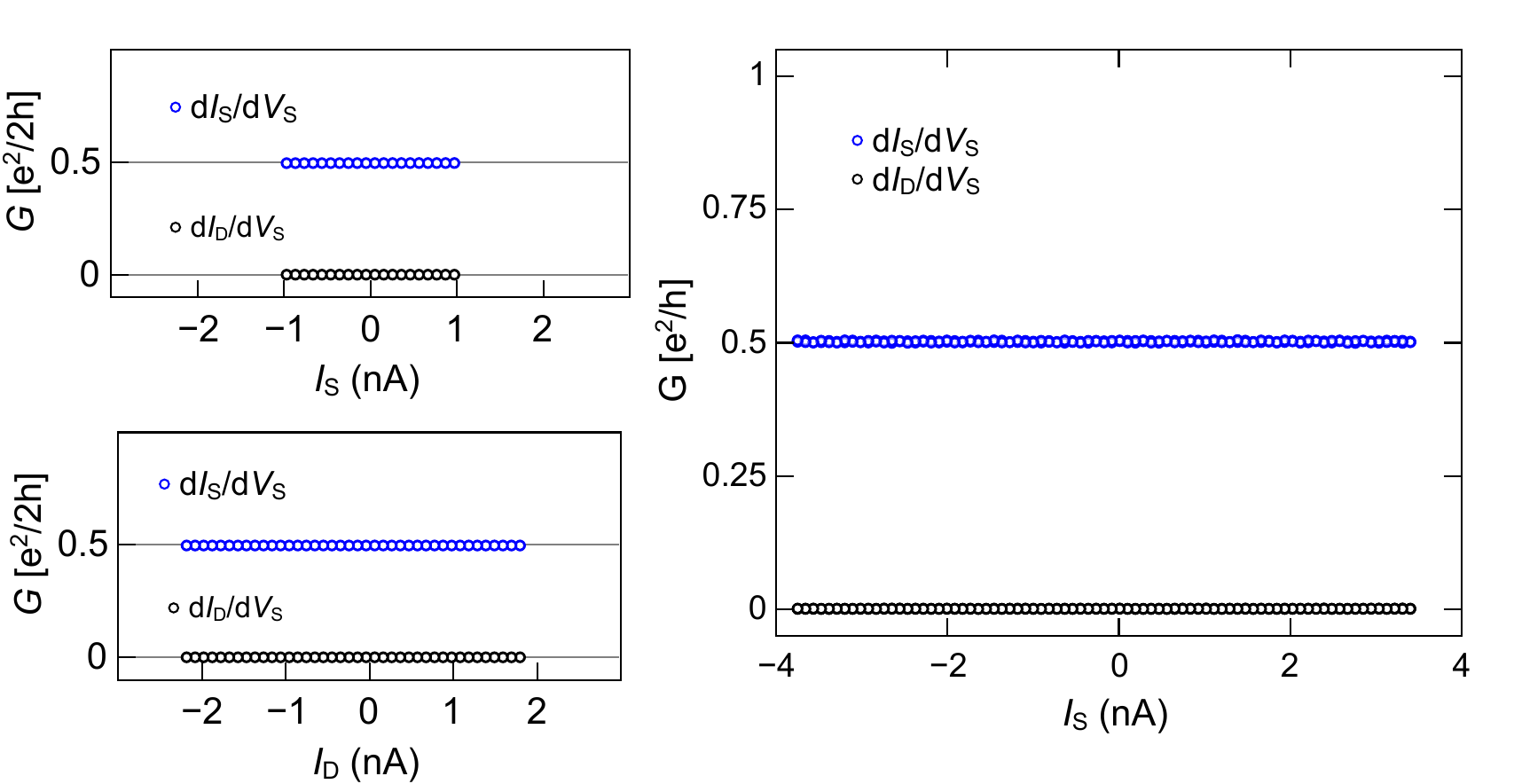}
\caption{\label{figsup-2term_nuc0_conductance} Reflected (blue) and transmitted (black) transconductance of the 2-terminal device versus applied dc current, corresponding to the $\nuc=0$ thermal transport data shown in main text Fig. 2.}
\end{figure}

\begin{figure}[ht]
\centering
\includegraphics[width=0.8\textwidth]{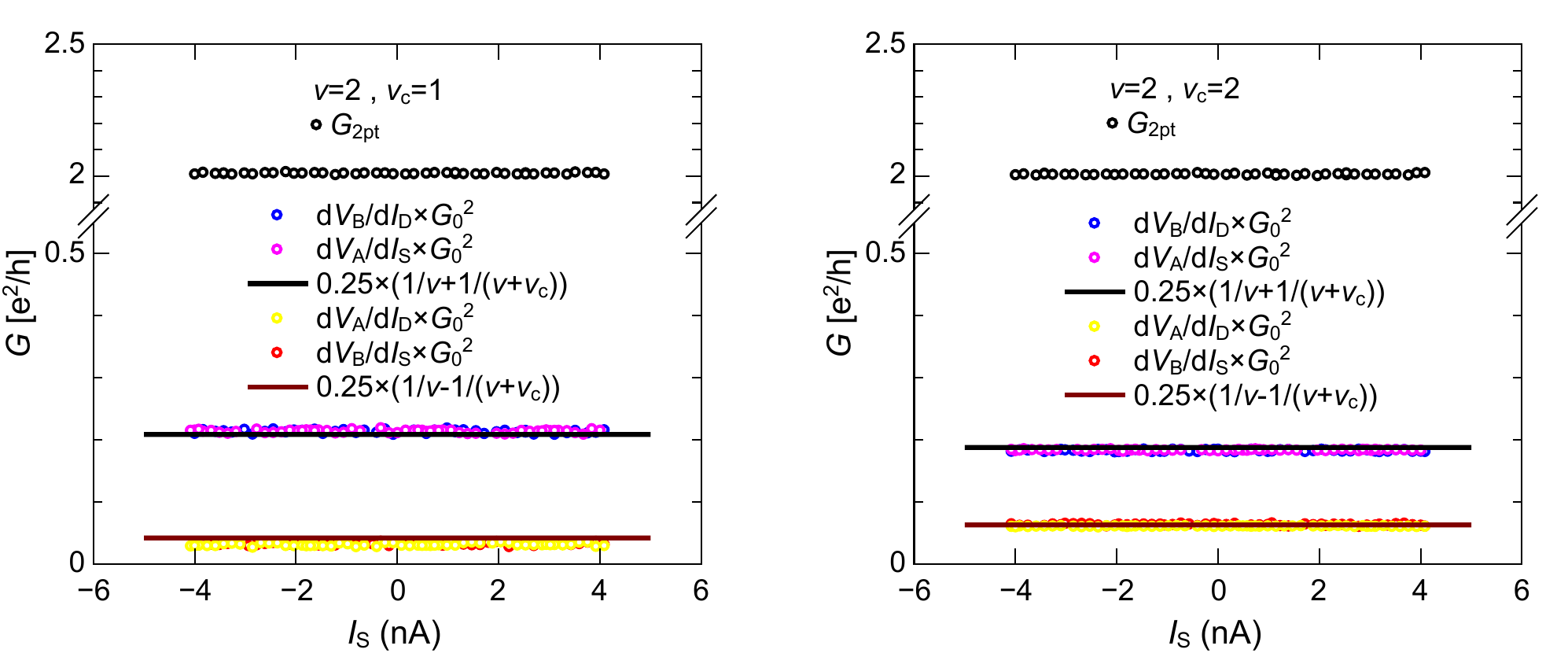}
\caption{\label{figsup-2term_cond2term} 2 point conductance (black), reflected (blue and magenta) and transmitted (yellow and red) transconductances of the 2-terminal device versus applied dc current for $\nuc=1$ (left) and $\nuc=2$ (right). The data correspond to the dataset shown in main text Figure 5a. The black and red lines are the expected values for the reflected and transmitted transconductances, respectively (see text).}
\end{figure}

Supplementary Fig.~\ref{figsup-2term_G_zerofield} shows the 2 point resistance of the two terminal device, measured from one of the conductance measurement terminals, versus the back gate voltage, at zero magnetic field and $T=60~$K. It shows the onset of satellite peaks at finite carrier density, indicating that the graphene flake is aligned with one of its encapsulating hBN crystals. From the approximate position of the satellite peak (a gate leakage prevented us from fully exploring the peaks, also leading to gate instabilities and noisy signals, see \textit{e.g.} the dip at zero gate voltage), we infer an alignment angle lower than 1 degree. This alignment leads to the development of a gap at the charge neutrality point, apperaing clearly when measuring the conductance across the central region of the device at low temperature. Supplementary Figure~\ref{figsup-2term_Fandiag} shows the transmitted transconductance across the $\nuc$ region as a function of the magnetic fielmd and the top gate voltage. An insulating state is clearly present even at zero magnetic field, indicating a broken sublattice symmetry. From the value of the angle, as well as previous works~\cite{Zibrov2018}, we estimate that gap to be about $30-40~$meV.

In the heat transport  experiments described in the main text, we simultaneously measure the noise (yielding the respective temperatures of the source and the detector) and the conductances, particularly making sure that no electrical current flows through the $\nuc=0$ region. This is exemplified in Supplementary Fig.~\ref{figsup-2term_nuc0_conductance}, showing the transmitted and reflected transconductances corresponding to the data shown in main text Figure 2, as a function of the dc current. In all cases, the reflected transconductance is quantized to $0.5e^2/h$ (reflecting that the floating electrodes evenly split the electrical current in two), while the transmitted transconductance is zero. Conversely, Supplementary Fig.~\ref{figsup-2term_cond2term} shows that for nonzero $\nuc$, the transmitted transconductance is finite, and both it and the reflected transconductance match their expected value computed above: $G_{R/T}=\frac{G_0}{4}(\frac{1}{\nu}\pm\frac{1}{\nu+\nuc})$.

\subsection{Heat Corbino device}

\begin{figure}[ht]
\centering
\includegraphics[width=0.8\textwidth]{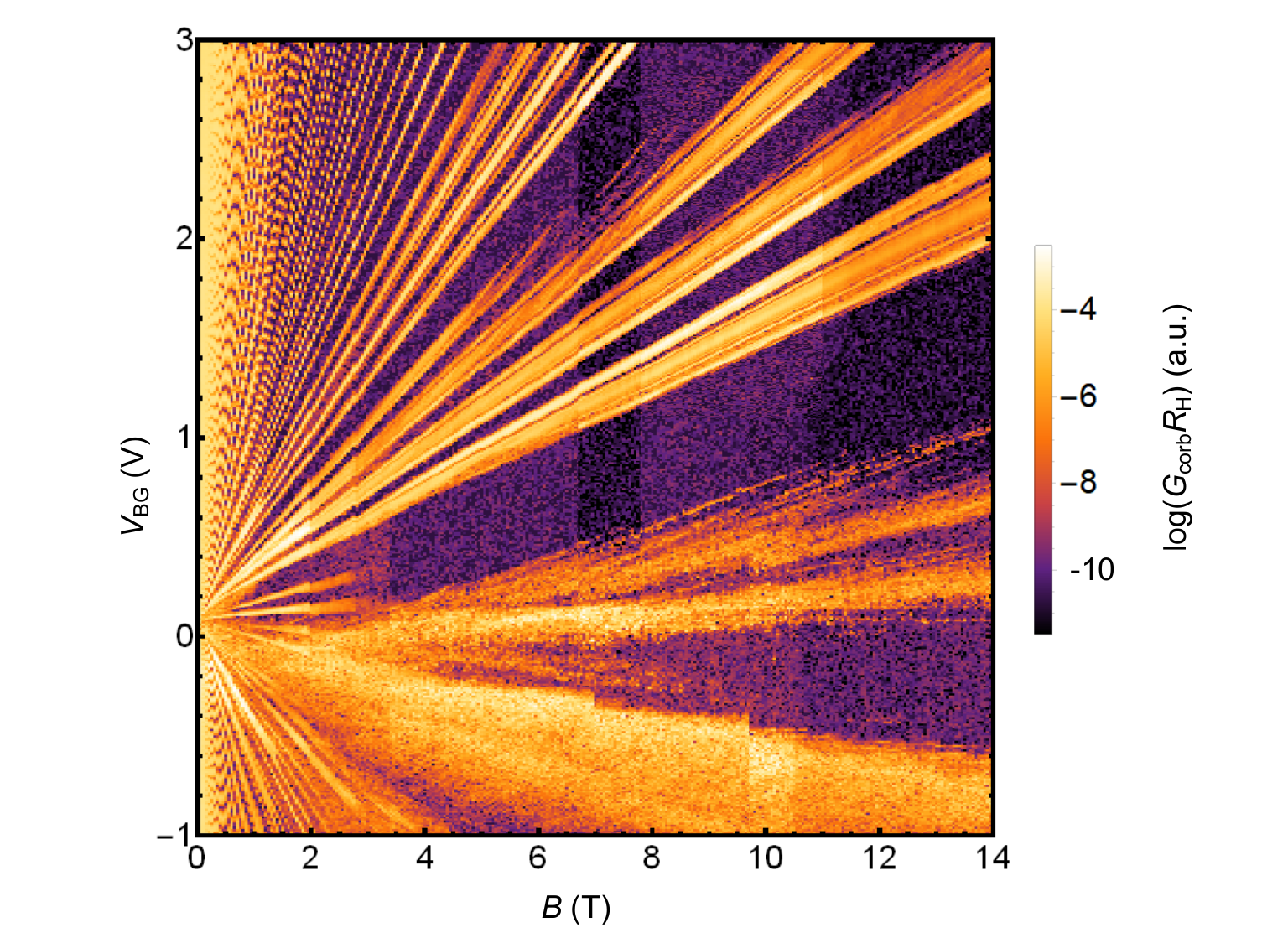}
\caption{\label{figsup-Corb_FanDiagram} Measured bulk electrical conductance of the bottom graphene of the Corbino heat transport device, versus back gate voltage and magnetic field, at $T\approx20~$mK.}
\end{figure}

\begin{figure}[ht]
\centering
\includegraphics[width=0.4\textwidth]{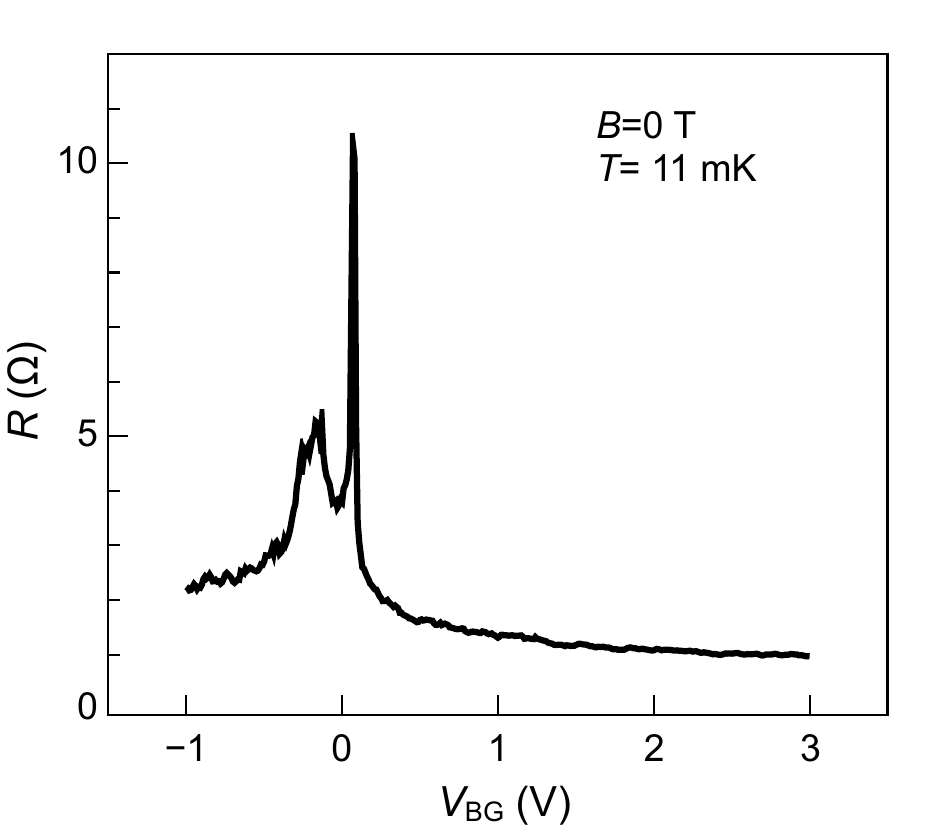}
\caption{\label{figsup-Corb_CNP} Measurement of the resistance of the bottom graphene of the Corbino device as a function of the back gate voltage. The additional peak at negative gate voltage corresponds to a local doping near the contact.}
\end{figure}

\begin{figure}[ht]
\centering
\includegraphics[width=0.4\textwidth]{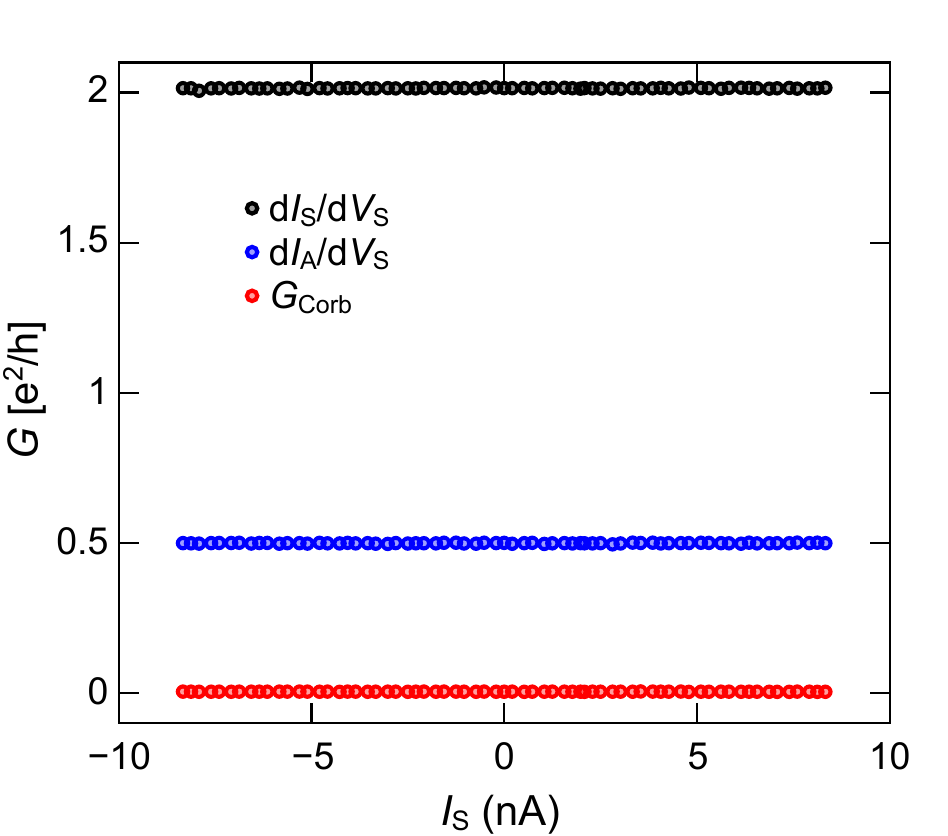}
\caption{\label{figsup-Corb_G_nutop2_nubot0_large} Measured top graphene 2 point conductance (black) and reflected transconductance (blue), as well as bottom graphene bulk conductance (red) in the Corbino heat transport device versus dc current. The data corresponds to the dataset shown in main text Figure 3a, with $B=7~$T, $\nu=2$ and $\nuc=0$.}
\end{figure}

\begin{figure}[ht]
\centering
\includegraphics[width=0.4\textwidth]{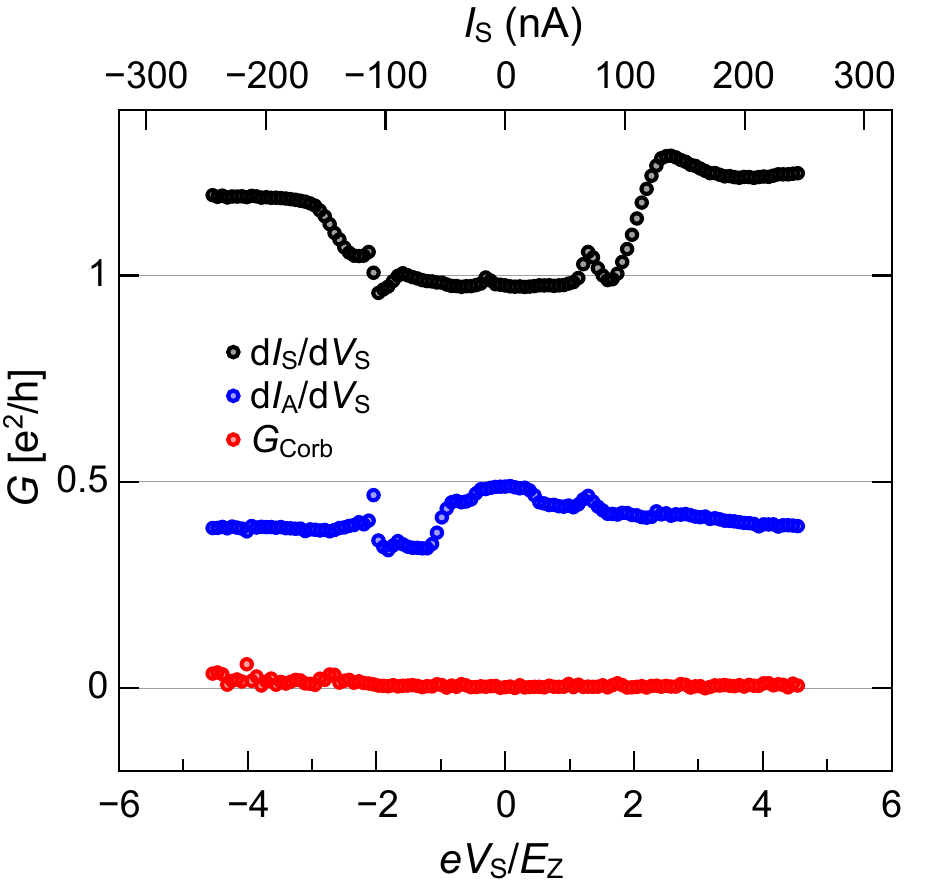}
\caption{\label{figsup-Corb_G_magnons} Measured top graphene 2 point conductance (black) and reflected transconductance (blue), as well as bottom graphene bulk conductance (red) in the Corbino heat transport device versus dc current. The data corresponds to the dataset shown in main text Figure 4, with $B=12~$T, $\nu=1$ and $\nuc=0$. The bottom X-axis shows the converted dc current (indicated in the top x-axis) into the dc voltage at the current feed contact of the device in units of the Zeeman energy at $B=12~$T.}
\end{figure}

Supplementary Fig.~\ref{figsup-Corb_FanDiagram} shows a measurement of the bulk electrical conductance of the bottom graphene in the heat Corbino device as a function of the back gate voltage and the magnetic field. $\Gcorb$ correspond to the transconductance between the dc current feed of the top graphene and the contact connected to the edge of the bottom graphene, and is normalized by the Hall resistance of the top graphene to compensate for changes in the filling factor of the top part as the magnetic field is swept. The absence of a zero magnetic field gap at the charge neutrality and of fractal features demonstrates that the bottom graphene is not aligned with its encapsulating hBN. The noisy features at negative back gate voltage are due to lock-in saturation caused by the large ac voltages developing on the bottom graphene edge contact at $\nuc=0$ as we simultaneously measure the two point edge conductance. These features are not exactly aligned with the finite Corbino conductance lines stemming from the charge neutrality point, indicating that the local doping near the edge contact is slightly different from that of the bulk. This appears clearly in the measurement of the electrical resistance of the bottom graphene at low temperature and zero magnetic field plotted in Supplementary Figure~\ref{figsup-Corb_CNP}, showing a very sharp peak at the charge neutrality point that stems from the bulk of the sample, and a second, broader one at negative gate voltage, corresponding to local doping in the vicinity of the edge contact. Note that despite its sharpness, the main charge neutrality point peak does not show signatures of a gap opening, indicating that the graphene flake is fully misaligned with both hBN encapsulating crystals.

As for the two terminal device, we simultaneously measure the electrical conductance while performing the thermal transport measurements. The conductance measurements corresponding to the data in main text Figure 3a are shown in Supplementary Figure~\ref{figsup-Corb_G_nutop2_nubot0_large}, demonstrating an absence electrical current flowing through the bulk of the bottom graphene while the electrical conductance of the top graphene is properly quantized. Similar measurements for the large dc bias data in main text Figure 4 are plotted in Supplementary Figure~\ref{figsup-Corb_G_magnons}, with significant deviations in the conductance quantization in the top graphene at $\IS>50~$nA. At $\nu=1$, this corresponds to dc voltages developing in the top graphene comparable to the Zeeman energy (lower X-axis), which leads to the emission and absorption of magnons in the ferromagnetic bulk of the top graphene~\cite{Wei2018}. Note that the Corbino conductance essentially remains zero, again demonstrating the absence of bulk electrical transport in the bottom graphene.

\section{Noise measurements}

\begin{figure}[h!]
    \centering
    \includegraphics[width=0.7\textwidth]{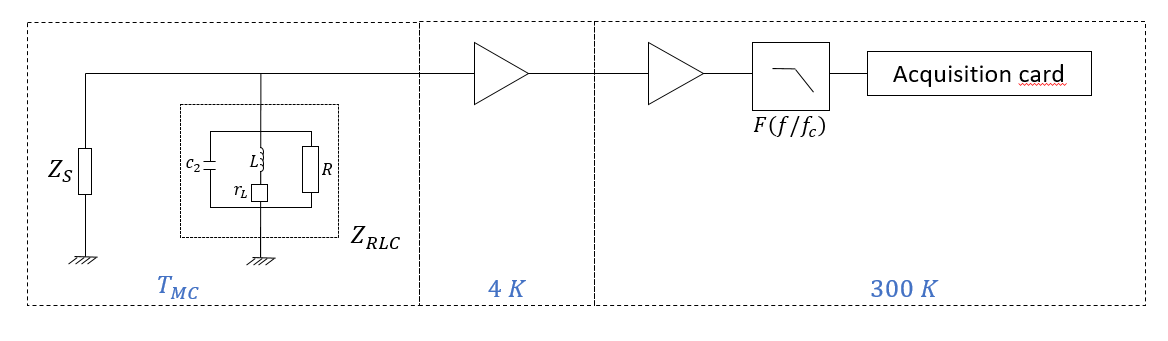}
    \caption{Noise measurement circuit from the sample to the acquisition card. The resonator is made with a capacitor $c_2$, an inductor $L$ with an effective resistance $r_L$ and a resistance $R$ which represents the losses on the circuit. }
    \label{fig:calib_circuit}
\end{figure}

\subsection{Calibration}

\subsubsection{Temperature calibration}

The noise which is measured is the following one :

\begin{equation}
        S^{meas}_{th,v}=G^2 \times \int_{BW} df F( \frac{f}{f_c} ) \left[ S_{v,amp}^2 +
    \left\lvert Z_{//} \right\rvert^2 \left( S^2_{i,amp} +4k_BTRe\left(\frac{1}{Z_{RLC}}\right)+ S^2_{i,sample}(T_S,\Tc=T_S) \right) \right]
    \label{eq:thnoise_meas}
\end{equation}

\begin{figure}[ht!]
    \centering
    \includegraphics[scale=0.70]{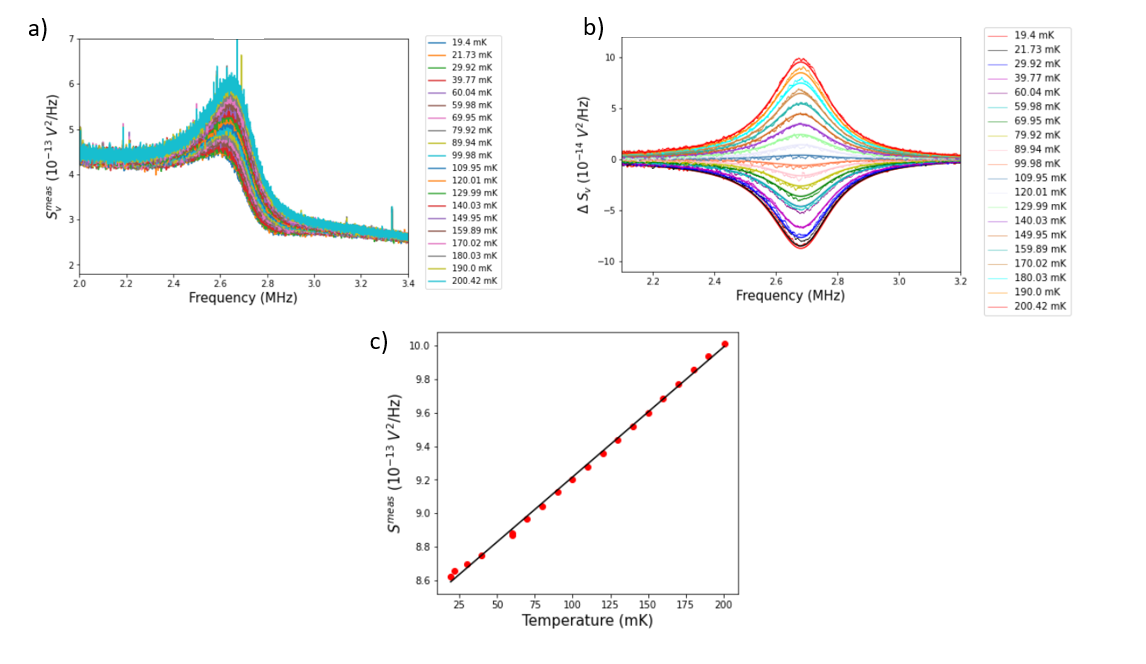}
    \caption{Temperature calibration at 14 T. a)b) Spectrum of the noise with the frequency for different temperature, where a) is direct noise measurement and b)is $\Delta S_v$ with the fitting function (straight lines). The figure c) is the $S^{meas}$ with the temperature of the fridge, where the linear fit is the black line.}
    \label{fig:calib_figures}
\end{figure}

Where $S^2_{i,amp}$ and $S^2_{v,amp}$, are the current and voltage noise of the amplifier, $4k_B T Re(\frac{1}{Z_{RLC}})$ the thermal noise of the resonator which is an LCR circuit, and $S^2_{i,sample}(T_S,\Tc)=3\kB T_S\nu e^2/h+\kB \Tc\nu e^2/h$ the current noise of the sample, where the metallic island can generally be at higher electron temperature $\Tc$ than the rest of the sample, at temperature $T_S$. The unknown are the amplifier noise, and the resonator noise. We determine these parameter with a temperature calibration where we measured the equilibrium noise for different fridge temperatures (measured using a ruthenium oxide thermometer) ranging between 10 and 200 mK, for various filling factors. Fig.~\ref{fig:calib_figures}a) shows typical raw spectra obtained from this calibration. We first remove the contribution of the temperature-independent terms by calculating the difference between each spectrum and the average of all spectra:
\begin{equation}
    \Delta S_v = S^{meas}_{th}-\left< S^{meas}_{th} \right>_T,
\end{equation}

yielding the curves shown in Fig.~\ref{fig:calib_figures}b), given by the equation, which assumes that all temperatures $T$, $T_S$ and $\Tc$ are equal:

\begin{equation}
    \Delta S^{meas}_v=G^2\int_{BW} F \left(\frac{f}{f_c}\right) 2k_B\Delta T \left\lvert Z_{//} \right\rvert^2 \left[ 2Re \left( \frac{1}{Z_{RLC}} \right) +\nu G_{el}  \right]
    \label{eq:calib_thermalnoise}
\end{equation}

The parameters of the LCR circuit will be found by fitting the above equation \ref{eq:calib_thermalnoise} from the measured noise for a fixed value of $\nu$. The voltage noise of the amplifier can be found from the intercept in temperature dependence of the integrated noise. If we look at the equation \ref{eq:thnoise_meas}, the noise of the amplifier doesn't depend on the temperature. Then, we can do a linear fit, and find  $S_{v,amp_A}\simeq 0.26 \text{ nV /} \sqrt{Hz}$ and $S_{v,amp_B}\simeq 0.28 \text{ nV /} \sqrt{Hz}$.
This calibration can be applied to the measurement, to analyse the current noise of the sample. The calibrated noise is the following :

\begin{multline}\\
    \Delta S = \frac{BW \Delta S^{meas}_v}{G^2 \int_{BW}df F(\frac{f}{f_c}) \left\lvert Z_{//}\right\rvert^2 } \\
    \\
    \Delta S = 2 k_B \nu G_{el} \Delta T \\
\end{multline}

\subsection{Noise thermometry}

We explain here the principle behind our using of auto and cross correlations to extract the temperature increase $\DeltaTS$ from spurious additional contributions in the heat Corbino geometry. We first begin by establishing the formula linking $\DeltaTS$ to the current noise spectral density $\Delta S$ flowing out of the sample. Following~\cite{Jezouin2013a}, a floating reservoir $\alpha$ at temperature $T_\alpha$ and voltage $V_\alpha$ emits in the $i$-th edge channel flowing out of it a current fluctuation given by:

\begin{equation}
    \delta I_i=\delta I_i^{T_\alpha}+G_0\delta V_\alpha
    \label{eq:currfluccontact}
\end{equation}

The first term $\delta I_i^{T_\alpha}$ corresponds to current fluctuations due to the finite temperature of the reservoir, with a spectral density $<(\delta I_i^{T_\alpha})^2>=2G_0 k_B T_\alpha$; the second term corresponds to fluctuations of the voltage of the floating reservoir. Importantly, in each of the edge channels, the first term is uncorrelated, $<\delta I_i^{T_\alpha}\delta I_j^{T_\alpha}>=\delta_{i,j}\times2G_0 k_B T_\alpha$, while the second is correlated, \textit{i.e.} it is equal at all times in all edge channels flowing out of the reservoir.
We adopt the following notations for the contacts: $\alpha=S$ for the central floating metallic island, $\alpha=A,B$ for the contacts connected to noise measurement lines with complex input impedances $Z_{A,B}$, and $\alpha=Ain,Bin$ for the current feed contacts upstream of the floating island on the $A,B$ side, the voltages of which are assumed to be without fluctuations. Assuming that we are in the integer QH regime with filling factor $\nu$, the current balances using Eq.~\ref{eq:currfluccontact} on contacts $A$, $B$ and $c$ read:

\begin{eqnarray}
\delta V_A (\nu G_0+1/Z_A)+\sum_i \delta I_i^{T_A}=\left(\sum_i \delta I_i^{T_S}\right)^A+\nu G_0\delta V_S\nonumber\\
\delta V_B (\nu G_0+1/Z_B)+\sum_i \delta I_i^{T_B}=\left(\sum_i \delta I_i^{T_S}\right)^B+\nu G_0\delta V_S\nonumber\\
\left(\sum_i \delta I_i^{T_S}\right)^A+\left(\sum_i \delta I_i^{T_S}\right)^B+2\nu G_0\delta V_S=\sum_i \delta I_i^{T_{Ain}}+\sum_i \delta I_i^{T_{Bin}}
    \label{eq:currflucbal}
\end{eqnarray}

The thermal noise of the measurement impedances $Z_{A,B}$ is neglected here for simplicity, and $\left(\sum_i \delta I_i^{T_S}\right)^{A/B}$ corresponds to the sum of the thermal current fluctuations flowing from the metallic island to contacts $A/B$. We combine these equations to express the voltage fluctuations $\delta V_A$ and $\delta V_A$ as a function of all other current fluctuations:

\begin{eqnarray}
\delta V_A (\nu G_0+1/Z_A)=-\sum_i \delta I_i^{T_A}-\frac{1}{2}\left(\sum_i \delta I_i^{T_S}\right)^A-\frac{1}{2}\left(\sum_i \delta I_i^{T_S}\right)^B+\frac{1}{2}\sum_i \delta I_i^{T_{Ain}}+\frac{1}{2}\sum_i \delta I_i^{T_{Bin}}\nonumber\\
\delta V_B (\nu G_0+1/Z_B)=-\sum_i \delta I_i^{T_B}-\frac{1}{2}\left(\sum_i \delta I_i^{T_S}\right)^A+\frac{1}{2}\left(\sum_i \delta I_i^{T_S}\right)^B+\frac{1}{2}\sum_i \delta I_i^{T_{Bin}}+\frac{1}{2}\sum_i \delta I_i^{T_{Ain}}
    \label{eq:volflucAB}
\end{eqnarray}

One can thus see right away that the terms containing thermal fluctuations of the metallic islands are anti-correlated, while the thermal fluctuations of $A/B$ are uncorrelated and those of $Ain/Bin$ are positively correlated. Assuming first that $T_A=T_B=T_{Ain}=T_{Bin}=T_0$ (that is, only the metallic island heats up while all other contacts stay at base electron temperature), we can calculate the auto and crosscorrelated voltage noise spectra:

\begin{eqnarray}
    <(\delta V_A)^2>=\frac{1}{|\nu G_0+1/Z_A|^2}\left[ 3\nu G_0 \kB T_0 + \nu G_0 \kB T_S \right]\nonumber\\
    <(\delta V_B)^2>=\frac{1}{|\nu G_0+1/Z_B|^2}\left[ 3\nu G_0 \kB T_0 +\nu G_0 \kB T_S\right]\nonumber\\
    <\delta V_A(\delta V_B)^\ast> =\frac{1}{(\nu G_0+1/Z_A)(\nu G_0+1/Z_B)^\ast}\left[ \nu G_0 \kB T_0 - \nu G_0 \kB T_S \right]
    \label{eq:volcorrAB}
\end{eqnarray}

This can finally be expressed as a function of the excess thermal current noise $\Delta S=\nu G_0 \kB (\TS-T_0)$:

\begin{eqnarray}
    <(\delta V_A)^2> =\frac{1}{|\nu G_0+1/Z_A|^2}\left[ 4\nu G_0 \kB T_0 + \Delta S\right]\nonumber\\
    <(\delta V_B)^2>=\frac{1}{|\nu G_0+1/Z_B|^2}\left[ 4\nu G_0 \kB T_0 + \Delta S\right]\nonumber\\
    <\delta V_A(\delta V_B)^\ast> =-\frac{1}{(\nu G_0+1/Z_A)(\nu G_0+1/Z_B)^\ast}\times\Delta S
    \label{eq:volcorrABdeltaS}
\end{eqnarray}

Thus, an increase of the metallic island electron temperature $\DeltaTS$ leads to positive autocorrelations, and negative crosscorrelation. Following \cite{Sivre2019}, we rely on this to separate the contribution of thermal noise due to the increase of the metallic island's electron temperature from additional spurious sources of noise. These sources can have three origins: current fluctuations generated upstream of the island, after the island, and at the island interface. In the previous calculations, the first two cased can be encompassed by introducing effective temperatures: $T_{Ain/Bin}$ for the upstream noise, and $T_{A/B}$ for the noise downstream of the island. Recalling the above equations, this leads to:

\begin{eqnarray}
    <(\delta V_A)^2>=\frac{1}{|\nu G_0+1/Z_A|^2}\left[ 2\nu G_0 \kB T_A + \nu G_0 \kB T_S +\frac{1}{2}\nu G_0 \kB T_{Ain} + \frac{1}{2}\nu G_0 \kB T_{Bin}\right]\nonumber\\
    <(\delta V_B)^2>=\frac{1}{|\nu G_0+1/Z_B|^2}\left[ 2\nu G_0 \kB T_B + \nu G_0 \kB T_S +\frac{1}{2}\nu G_0 \kB T_{Bin} + \frac{1}{2}\nu G_0 \kB T_{Ain}\right]\nonumber\\
    <\delta V_A(\delta V_B)^\ast> =\frac{1}{(\nu G_0+1/Z_A)(\nu G_0+1/Z_B)^\ast}\left[- \nu G_0 \kB T_S +\frac{1}{2}\nu G_0 \kB T_{Bin} + \frac{1}{2}\nu G_0 \kB T_{Ain}\right]
    \label{eq:volcorrABdifftemp}
\end{eqnarray}

An additional upstream noise increases both auto and crosscorrelations by the same amount, which can be subtracted by computing the difference between auto and crosscorrelation. Conversely, noise downstream of the island will lead to different autocorrelations. In that case, it becomes problematic to extract the thermal contribution, as the added noise cannot be easily subtracted. We thus discard the data as no reliable heat transport analysis can be performed.

The case of current fluctuations generated at the island interface requires additional analysis. We present here a simple model based on a single edge channel for simplicity. The imperfect interface is modelled by a scatterer with transmission $\tau$ inserted before the island. We assume here that only one interface (the one connected to the A side) is imperfect, such that the scatterer reflects the current stemming from contact $Ain$ to contact $A$ with a coefficient $1-\tau$. The scatterer generates additional shot noise noted $\delta I^\ast$. Current conservation implies that the fluctuation thusly generated on either side of the scatterer is opposite: by convention, we write $-\delta I^\ast$ the current fluctuation emitted in the edge channel flowing to $A$, and $+\delta I^\ast$ the fluctuation emitted in the edge channel going to the island. This changes the current balances shown in Eq.~\ref{eq:currflucbal} to the following:

\begin{eqnarray}
\delta V_A (G_0+1/Z_A)+\delta I^{T_A}=-\delta I^\ast+(1-\tau)\delta I^{T_Ain}+\tau\left( \delta I^{T_S}\right)^A+\tau G_0\delta V_S \nonumber\\
\delta V_B (G_0+1/Z_B)+ \delta I^{T_B}=\left( \delta I^{T_S}\right)^B+G_0\delta V_S\nonumber\\
(1+\tau)G_0\delta V_S+\tau\left(\delta I^{T_S}\right)^A+\left(\delta I^{T_S}\right)^B=\delta I^\ast+\tau\delta I^{T_{Ain}}+\delta I^{T_{Bin}}
    \label{eq:currflucbalshotnoise}
\end{eqnarray}

Expressing, as above, $\delta V_{A/B}$ as a function of all current fluctuations yields (noting $F(\tau)=\tau/(1+\tau)$):

\begin{eqnarray}
\delta V_A (G_0+1/Z_A)=-\delta I^{T_A}-\frac{F(\tau)}{\tau}\delta I^\ast+F(\tau)\left(\delta I^{T_S}\right)^A-F(\tau)\left( \delta I^{T_S}\right)^B+\left[ 1-\tau+F(\tau)\right]\delta I^{T_{Ain}}+F(\tau) \delta I^{T_{Bin}}\nonumber\\
\delta V_B (G_0+1/Z_B)=-\delta I^{T_B}+\frac{F(\tau)}{\tau}\delta I^\ast-F(\tau)\left(\delta I^{T_S}\right)^A+\frac{F(\tau)}{\tau}\left( \delta I^{T_S}\right)^B+F(\tau)\delta I^{T_{Ain}}+\frac{F(\tau)}{\tau} \delta I^{T_{Bin}}
    \label{eq:volflucABshotnoise}
\end{eqnarray}

Thus, the contribution of the shot noise $<(\delta I^\ast)^2>$ is the same for auto and crosscorrelations, with a negative sign for the cross correlations. While this is quite similar to the contribution of the thermal noise for a perfect interface, it turns out that this same thermal noise has now different contributions in the two autocorrelations. This can be intuitively understood by the fact that the side with the imperfect interface sees less noise stemming from the island. In the end, by using the equal autocorrelation criterion mentioned above, one can again make sure that the noise measured only stems from the increase island temperature $T_S$.

Note that a poor interface at the floating island will also impact the conductance measurements, with the reflected differential transconductance being larger than the transmitted one. Checking that the transconductance are equal and quantized is thus crucial in these measurements.
Interestingly, a bad transparency of the injection contact upstream of the floating island will not necessarily appear in the conductance measurement, as the sample is current biased (an imperfect transparency will reduce the voltage drop on the injection contact, and add noise, but won’t change the amount of current fed flowing from this contact). Finally, our sample design is such that the floating island has a larger interface length than the other contacts (which is usually the opposite in GaAs thermal transport experiments). Thus, poor interfaces are more likely to stem from these contacts.

\begin{figure*}[h!]
\centering
\includegraphics[width=0.75\textwidth]{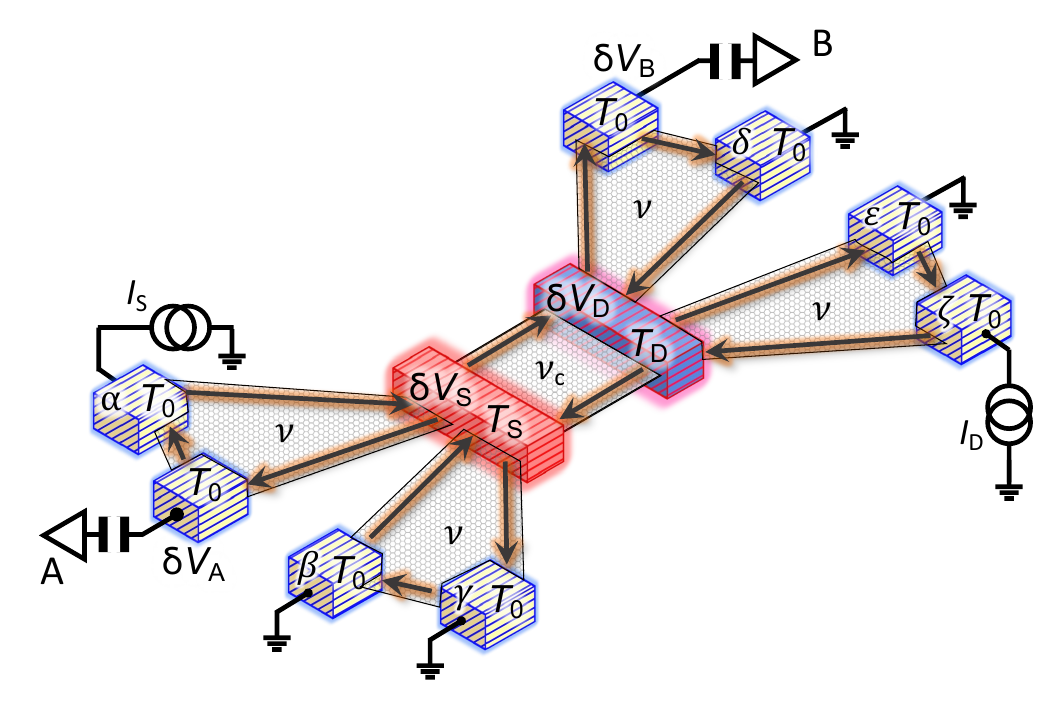}
\caption{\label{figsup-2term-sample-calc} Schematic representation of the 2 terminal sample.}
\end{figure*}

\subsection{Noise thermometry in the 2-terminal geometry}

For non-zero $\nuc$, the source and detector temperature are extracted from the auto- and cross-correlated noise measurement in A and B according to the following calculation, which extends the formalism presented in the previous section to the 2 terminal geometry. The current fluctuations balances on noise measurements contacts A and B, source, and detector read:

\begin{eqnarray}
\delta V_A (\nu G_0+1/Z_A)+\sum_i \delta I_i^{T_A}=\left(\sum_i \delta I_i^{T_S}\right)^A+\nu G_0\delta V_S\nonumber\\
\delta V_B (\nu G_0+1/Z_B)+\sum_i \delta I_i^{T_B}=\left(\sum_i \delta I_i^{T_D}\right)^B+\nu G_0\delta V_D\nonumber\\
\left(\sum_i \delta I_i^{T_S}\right)^A+\left(\sum_i \delta I_i^{T_S}\right)^\gamma+\left(\sum_i \delta I_i^{T_S}\right)^D+2\nu G_0\delta V_S+\nuc G_0\delta V_S=\sum_i \delta I_i^{T_{\alpha}}+\sum_i \delta I_i^{T_{\beta}}+\sum_i \delta I_i^{T_D}+\nuc G_0\delta V_D\nonumber\\
\left(\sum_i \delta I_i^{T_D}\right)^B+\left(\sum_i \delta I_i^{T_D}\right)^\epsilon+\left(\sum_i \delta I_i^{T_D}\right)^S+2\nu G_0\delta V_D+\nuc G_0\delta V_D=\sum_i \delta I_i^{T_{\delta}}+\sum_i \delta I_i^{T_{\zeta}}+\sum_i \delta I_i^{T_S}+\nuc G_0\delta V_S,
    \label{eq:currflucbal2term}
\end{eqnarray}

with $\alpha,\beta,\gamma,\delta,\epsilon,\zeta$ denoting the cold electrodes (shown in Supplementary Fig.~\ref{figsup-2term-sample-calc}), assumed to be at fixed potential and base electron temperature $T_0$. For simplicity, the thermal noise of the noise measurement impedances $Z_{A,B}$ is (as in the previous section) neglected. The current fluctuations balances on A and B allow us to compute the auto- and cross-correlated noises $\Delta S_\mathrm{A}$, $\Delta S_\mathrm{B}$ and $\Delta S_\mathrm{AB}$:

\begin{eqnarray}
\Delta S_\mathrm{A}=<(\delta V_A)^2> |\nu G_0+1/Z_A|^2=2\nu \kB G_0(\Ts-T_0)+\nu^2 G_0^2<(\delta V_S)^2>\nonumber\\
\Delta S_\mathrm{B}=<(\delta V_B)^2> |\nu G_0+1/Z_B|^2=2\nu \kB G_0(\Td-T_0)+\nu^2 G_0^2<(\delta V_D)^2>\nonumber\\
\Delta S_\mathrm{AB}=<\delta V_A(\delta V_B)^\ast> (\nu G_0+1/Z_A)(\nu G_0+1/Z_B)^\ast=\nu^2 G_0^2<\delta V_S(\delta V_D)^\ast>.
    \label{eq:volcorrABdeltaS2term}
\end{eqnarray}

Because of the finite $\nuc$, the correlator $<\delta V_S(\delta V_D)^\ast>$ is non zero. $<(\delta V_S)^2>$, $<(\delta V_D)^2>$ and $<\delta V_S(\delta V_D)^\ast>$ are then obtained from the current fluctuations balances on source and detector:

\begin{eqnarray}
G_0^2[(2\nu+\nuc)^2-\nuc^2]\times[<(\delta V_S)^2>-<(\delta V_D)^2>]= 4(\nu+\nuc) \kB G_0(\Td-\Ts)\nonumber\\ 
G_0^2[(2\nu+\nuc)^2+\nuc^2]\times[<(\delta V_S)^2>+<(\delta V_D)^2>]-4\nuc(2\nu+\nuc)  G_0^2<\delta V_S(\delta V_D)^\ast>= 4\nu \kB G_0(2T_0-\Ts-\Td).
    \label{eq:volcorrSD2term}
\end{eqnarray}

Combining the  above equations yields the expressions for $\Delta T_{S,D}$ as a function of $\Delta S_\mathrm{A}$, $\Delta S_\mathrm{B}$ and $\Delta S_\mathrm{AB}$:

\begin{eqnarray}
\Delta \Ts=\frac{1}{4 \nu G_0 \kB}\left( \frac{(2\nu+\nuc)^2+\nuc^2}{2(\nu+\nuc)^2}[\Delta S_\mathrm{A}+\Delta S_\mathrm{B}]+2[\Delta S_\mathrm{A}-\Delta S_\mathrm{B}]-\frac{2\nuc(2\nu+\nuc)}{(\nu+\nuc)^2}\Delta S_\mathrm{AB}\right) \nonumber\\
\Delta \TD=\frac{1}{4 \nu G_0 \kB}\left( \frac{(2\nu+\nuc)^2+\nuc^2}{2(\nu+\nuc)^2}[\Delta S_\mathrm{A}+\Delta S_\mathrm{B}]+2[\Delta S_\mathrm{B}-\Delta S_\mathrm{A}]-\frac{2\nuc(2\nu+\nuc)}{(\nu+\nuc)^2}\Delta S_\mathrm{AB}\right).
    \label{eq:TSTD2term}
\end{eqnarray}

This illustrates that measuring the cross-correlated noise signals is mandatory in order to independently extract the detector and source temperatures.

As mentioned in the main text, the 2-terminal heat transport measurements were realized in presence of a finite dc current offset: as \textit{e.g.} $\Is$ was swept, a constant and large $\Id$ of a few nA was inadvertently kept flowing into the device. This leads to an increase in the base electron temperature $T_0$, that is technically different for the source and detector contacts. For the sake of simplicity, the data presented in the main text is analyzed with the same $T_0$ for both.

\subsection{Noise measurements in the heat Corbino device}

\begin{figure*}[h!]
\centering
\includegraphics[width=0.5\textwidth]{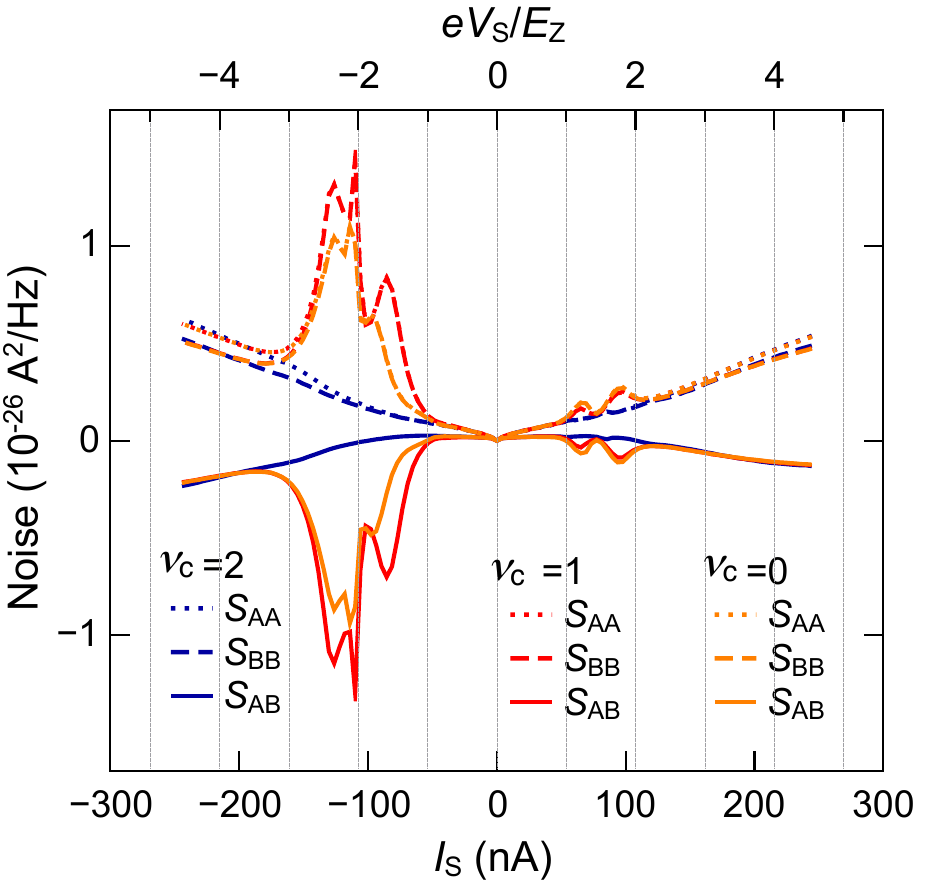}
\caption{\label{figsup-Corb_magnon_crossvsauto} Auto and cross correlated noise data corresponding to the data shown in main text Fig. 4. For clarity, the sign of the cross-correlated noise $S_\mathrm{AB}$ was inverted: most of its variations thus are actually positively correlated, indicating that their origin is not due to an increase in the source temperature.}
\end{figure*}

Supplementary Fig.~\ref{figsup-Corb_magnon_crossvsauto} shows the auto- and cross-correlated noise measurements corresponding to the data appearing in main text Fig.~4. The top X-axis displays the conversion from the dc current $\Is$ to the dc voltage developing on the sample in units of the Zeeman energy at 12~T. As in Supplementary Fig.~\ref{figsup-Corb_G_magnons}, as soon as the voltage exceeds the Zeeman energy, the noise signal strongly deviates, corresponding to the fact that magnons are emitted and absorbed in the top graphene set to $\nu=1$. These deviations are particularly strong for $\nuc=1$ and $\nuc=0$ at negative bias, such that an error of $1~\%$ on the calibration yields the very large error bars on the extracted $\Delta \Ts$ shown in main text Fig.~4. The origin between these large deviations are still under investigation.

\begin{figure*}[h!]
\centering
\includegraphics[width=0.45\textwidth]{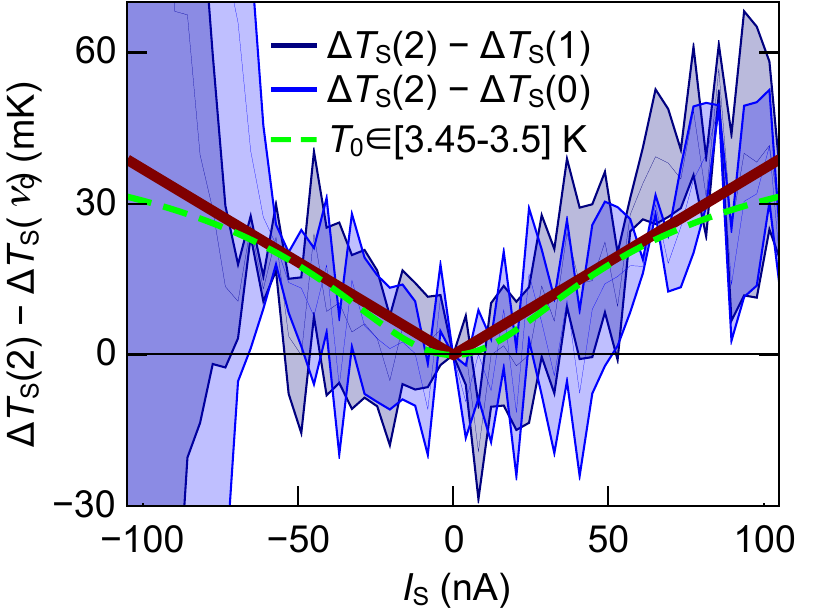}
\caption{\label{figsup-Corb_magnon_diffT0fit} Zoom on the temperature differences for $\vert\Is\vert<70~$nA in the heat Corbino data shown in main text Fig.~4. The dark red line corresponds to the additional thermal flow with $\kappa\sim 0.02 \kappa_0$, the green dashed line to a difference in base temperature between the $\nuc=2$ and the $\nuc=1,0$ data of about $3.45~$K, with no additional heat flow.}
\end{figure*}

Supplementary Fig.~\ref{figsup-Corb_magnon_diffT0fit} shows the differential temperature data corresponding to the inset of main text Fig.~4. Fitting the temperature difference with a purely electronic transport model assuming different base temperature yields a $T_0$ of about $3.45~$K, which is highly unlikely.

\section{2-terminal device 2}

We add here additional measurements on a second, non-aligned device in the 2-terminal geometry. The measurements were performed in the same refrigerator as the other devices, although with a different measurement setup (different cryogenic amplifiers and different RLC circuits with a resonance around $0.8~$MHz for the noise measurements, and a low-temperature DC voltage bias scheme realized with a $\sim300~\Omega$ shunt resistor in parallel to the sample on the dc current feed lines).

\subsection{Device fabrication and characterization}

\begin{figure*}[h!]
\centering
\includegraphics[width=0.75\textwidth]{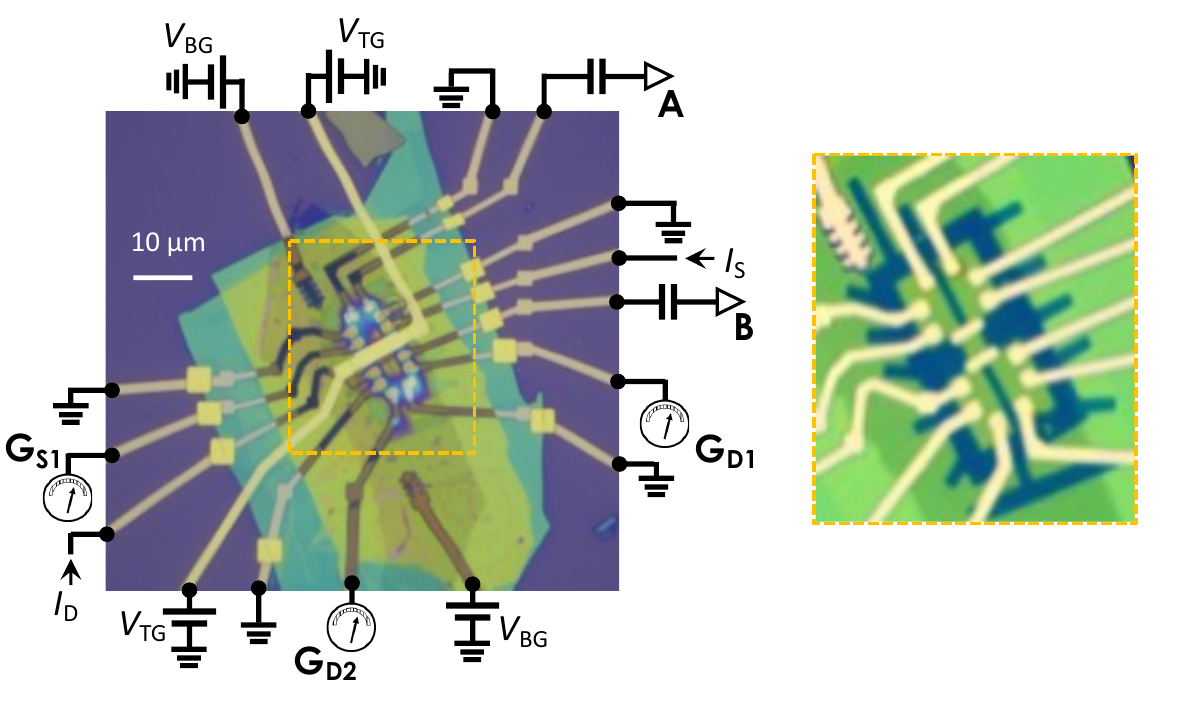}
\caption{\label{figsup-2termsample2} Optical micrograph of 2-terminal device 2, along with its wiring configuration. Inset on the right: zoom on the sample before depositing the graphite/hBN stack used for the top gate.}
\end{figure*}

\begin{figure*}[h!]
\centering
\includegraphics[width=0.45\textwidth]{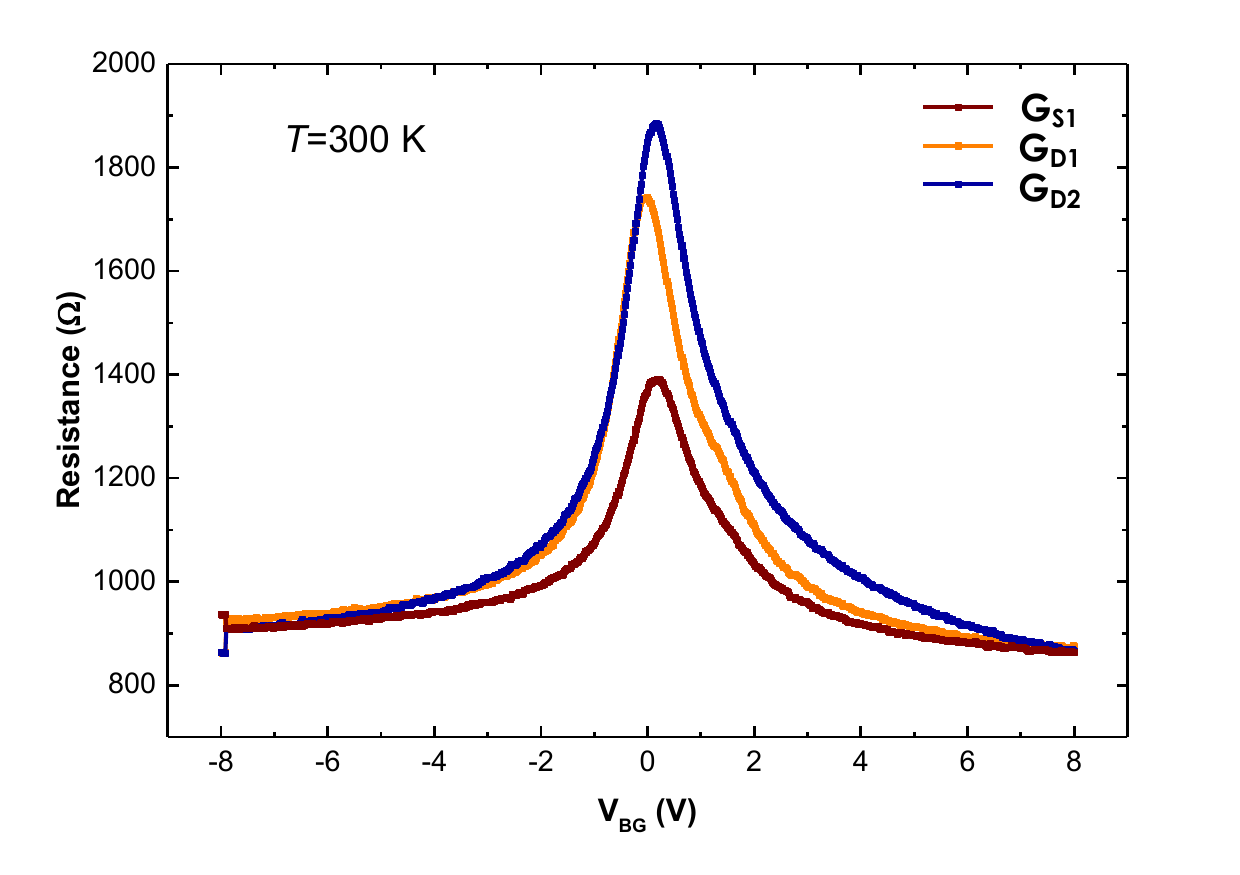}
\caption{\label{figsup-2termsample2-RvsVg300K} Two-point measurements of the resistance of 2-terminal device 2 at room temperature as a function of the back gate voltage. Measurements performed between each of the three conductance measurements contacts shown in Supplementary Figure~\ref{figsup-2termsample2} and one of the cold ground contacts.}
\end{figure*}

The geometry and fabrication process for this sample was similar to that of the first 2-terminal device (device 1). The top gate tuning the filling factor $\nuc$ was made by depositing a graphite/hBN heterostructure on top of the processed device, after which the graphite was etched to only cover the central region using a metallic mask.
$R(\Vbg)$ measurements at room temperature (see Supplementary Fig.~\ref{figsup-2termsample2-RvsVg300K}) showed no satellite peak, indicated that the graphene flake was not aligned with either top or bottom hBN. Low temperature conductance measurements, shown in Supplementary Fig.~\ref{figsup-2termsample2-RvsVgB=0} confirmed this observation with an absence of gap at the charge neutrality point. At finite magnetic field (Supplementary Fig.~\ref{figsup-2termsample2-RvsVgB=14T}), we observed that the outer regions is of substantially lower mobility than the central region, with no visible QH plateaus corresponding to $\nu=0$ and $\nu=1$ while they are well developed for $\nuc=0$ and $\nuc=1$.  We attribute this to the etching process of the top gate which is likely to have substantially damaged the surface of the hBN layers covering the graphene flake in the outer regions.

\begin{figure*}[h!]
\centering
\includegraphics[width=0.45\textwidth]{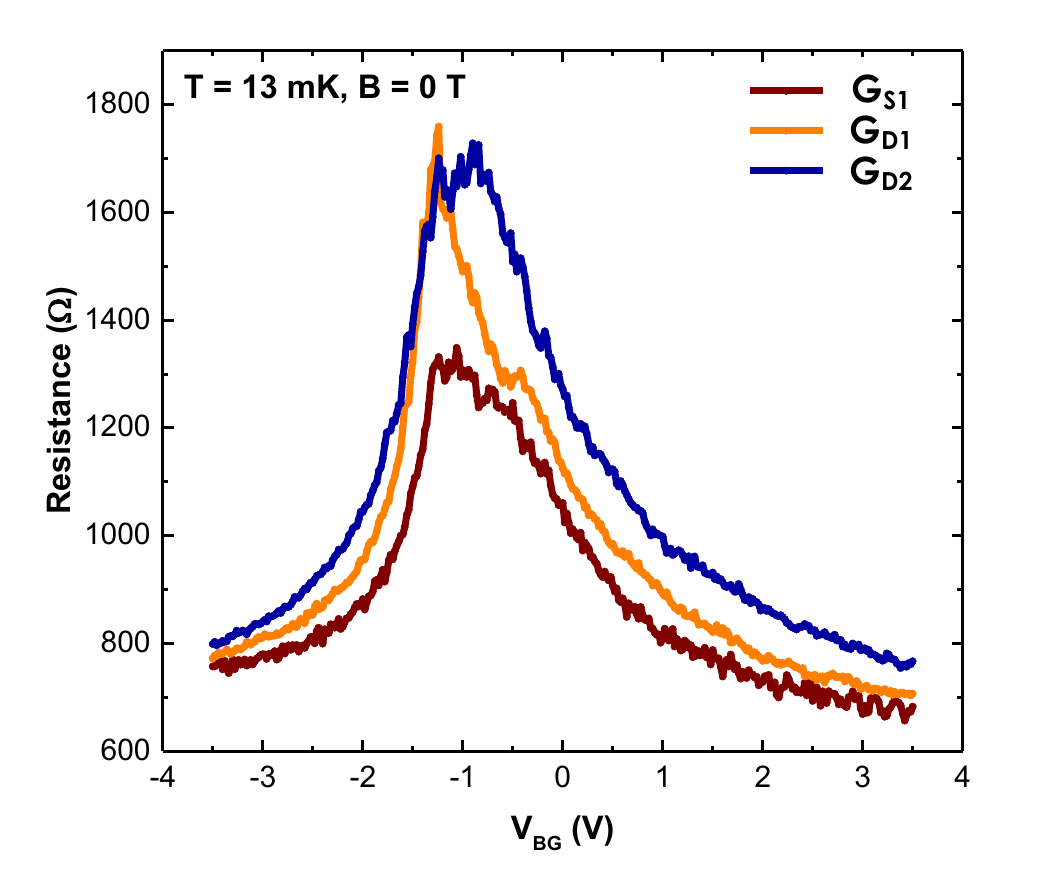}
\caption{\label{figsup-2termsample2-RvsVgB=0} Two-point measurements of the resistance of 2-terminal device 2 at $B=0~$T and $T=13~$mK as a function of the back gate voltage. Measurements performed between each of the three conductance measurements contacts shown in Supplementary Figure~\ref{figsup-2termsample2} and the cold ground.}
\end{figure*}

\begin{figure*}[h!]
\centering
\includegraphics[width=0.95\textwidth]{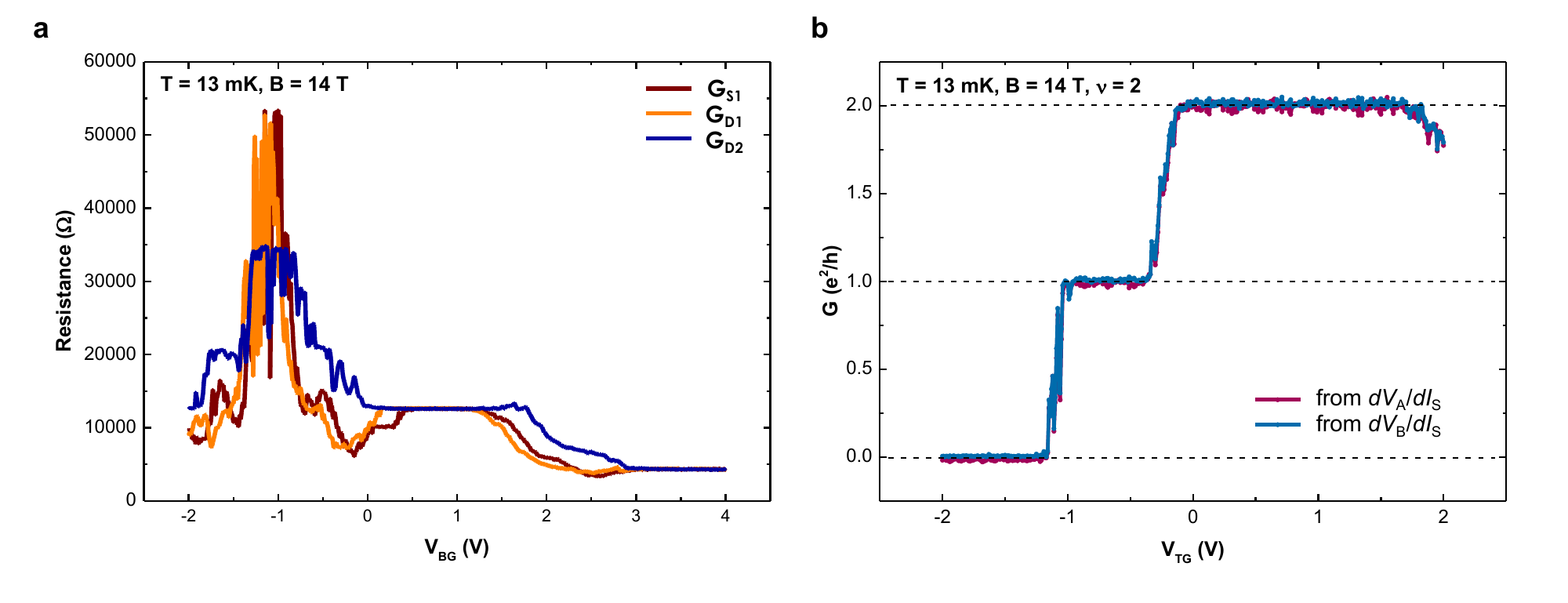}
\caption{\label{figsup-2termsample2-RvsVgB=14T} \textbf{a),} Measurement of the resistance of 2-terminal device 2 at $B=14~$T and $T=13~$mK as a function of the back gate voltage. The configuration is similar to that of the zero magnetic field measurements above. \textbf{b),} Measurements of the conductance of the central region as a function of the top gate voltage, for the filling factor of the outer region fixed to $\nu=2$ at $B=14~$T and $T=13~$mK. The conductance is extracted from the transconductance signals $G_0^2 dV_{A/B}/dI_S$ (see above).}
\end{figure*}

\begin{figure*}[h!]
\centering
\includegraphics[width=0.95\textwidth]{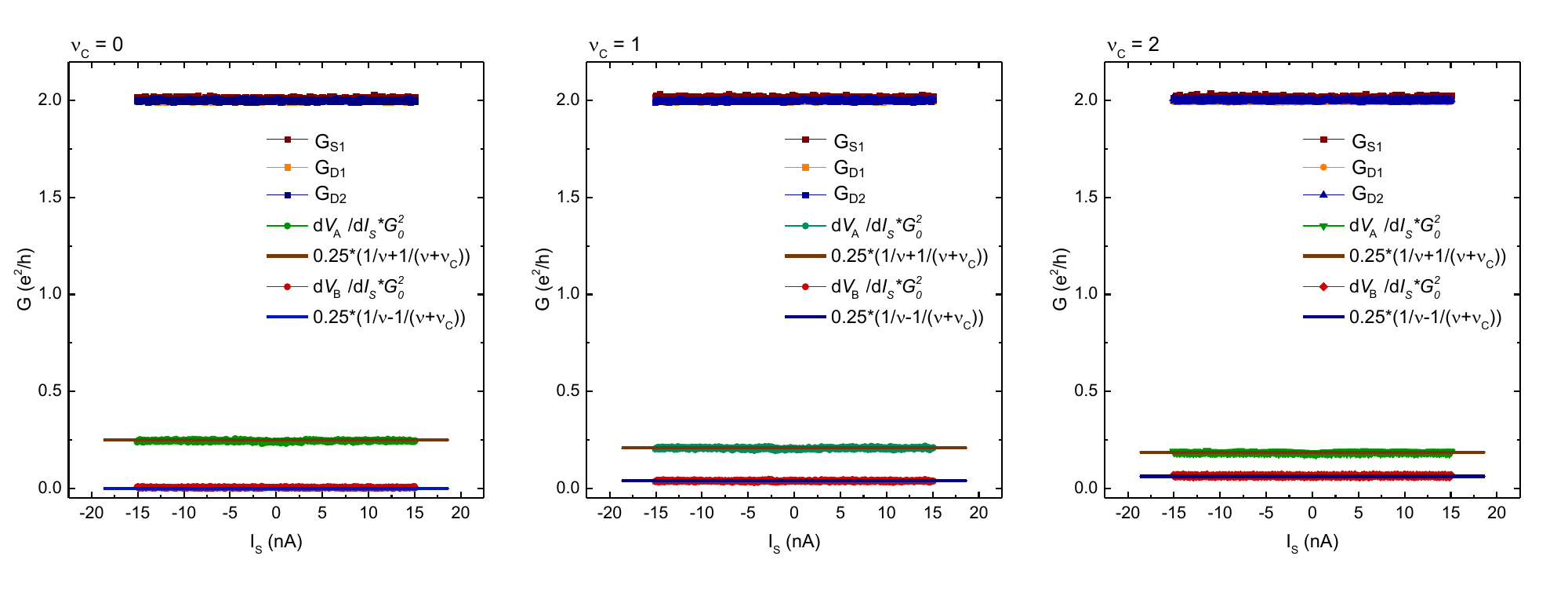}
\caption{\label{figsup-2termsample2-RvsIdc} 2-point conductances (squares) as well as reflected (green circles) and transmitted (red circles) transconductances (see above) measured as a function of $\Is$, for $\nuc=0$ (left), $\nuc=1$ (middle), and $\nuc=2$ (right), at $B=14~$T and $T\approx10~$mK.}
\end{figure*}

While performing the heat transport measurements shown in the next subsection, we simultaneously checked that all the source and detector had negligible contact resistor by performing transconductance measurement as a function of the current bias similar to those shown in Supplementary Fig.~\ref{figsup-2term_cond2term}. These measurements are shown in Supplementary Fig.~\ref{figsup-2termsample2-RvsIdc}, with an excellent agreement with the expected values of the transconductances, and no dc current bias dependence.

\subsection{Heat transport measurements}

\begin{figure*}[h!]
\centering
\includegraphics[width=0.85\textwidth]{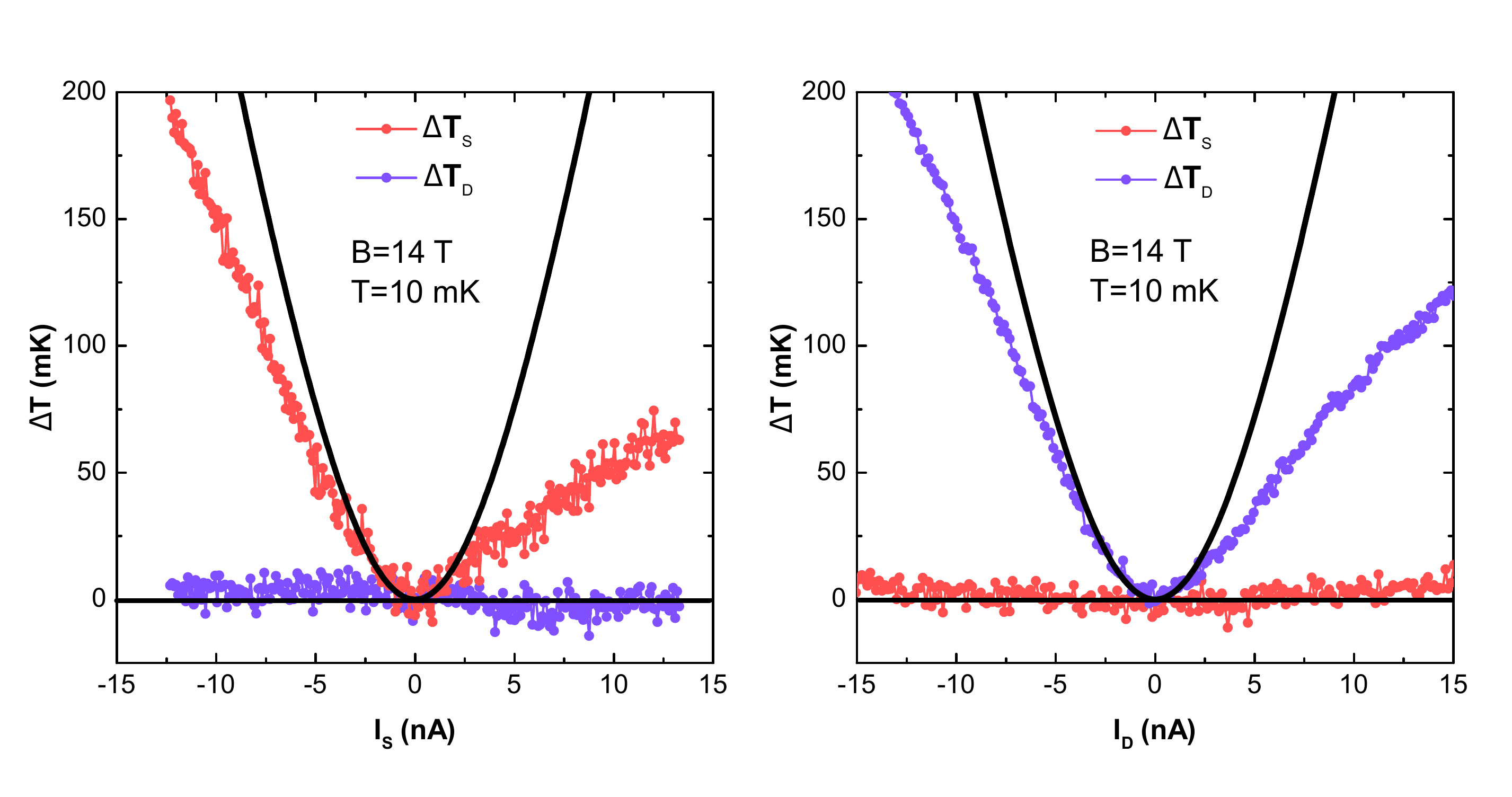}
\caption{\label{figsup-2termsample2-TvsI10mK} Temperature increase in the source $\Delta\Ts$ (red) and in the detector $\Delta\Td$ (blue) as a function of $\Is$ (left) and $\Id$ (right), measured at 14~T and 10~mK for $\nuc=0$. The averaging time per point is 80~s for the left panel, and 300~s for the right panel.}
\end{figure*}

\begin{figure*}[h!]
\centering
\includegraphics[width=0.85\textwidth]{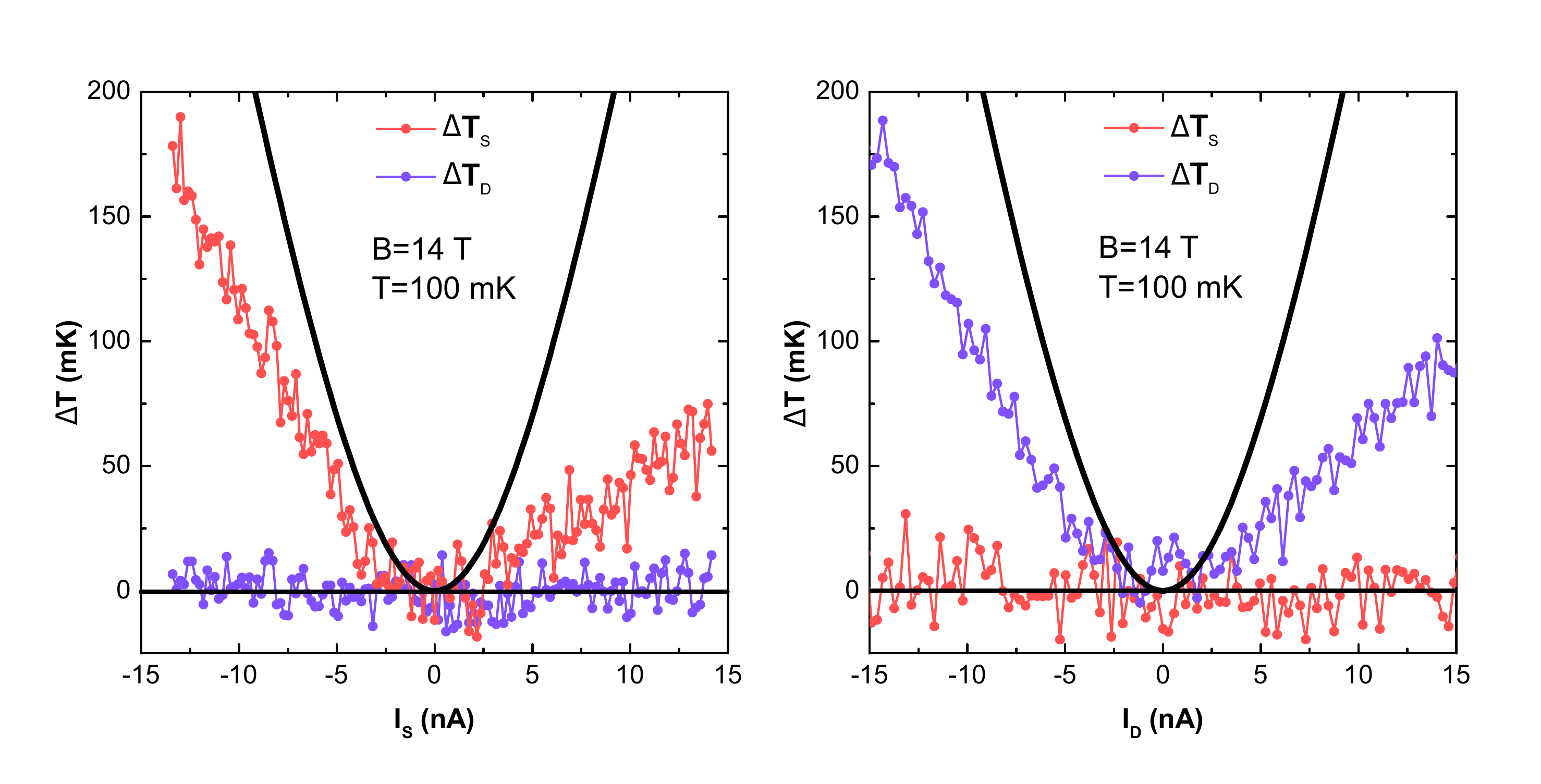}
\caption{\label{figsup-2termsample2-TvsI100mK} Temperature increase in the source $\Delta\Ts$ (red) and in the detector $\Delta\Td$ (blue) as a function of $\Is$ (left) and $\Id$ (right), measured at 14~T and 10~mK for $\nuc=0$. The averaging time per point is 30~s for both panels.}
\end{figure*}

\begin{figure*}[h!]
\centering
\includegraphics[width=0.75\textwidth]{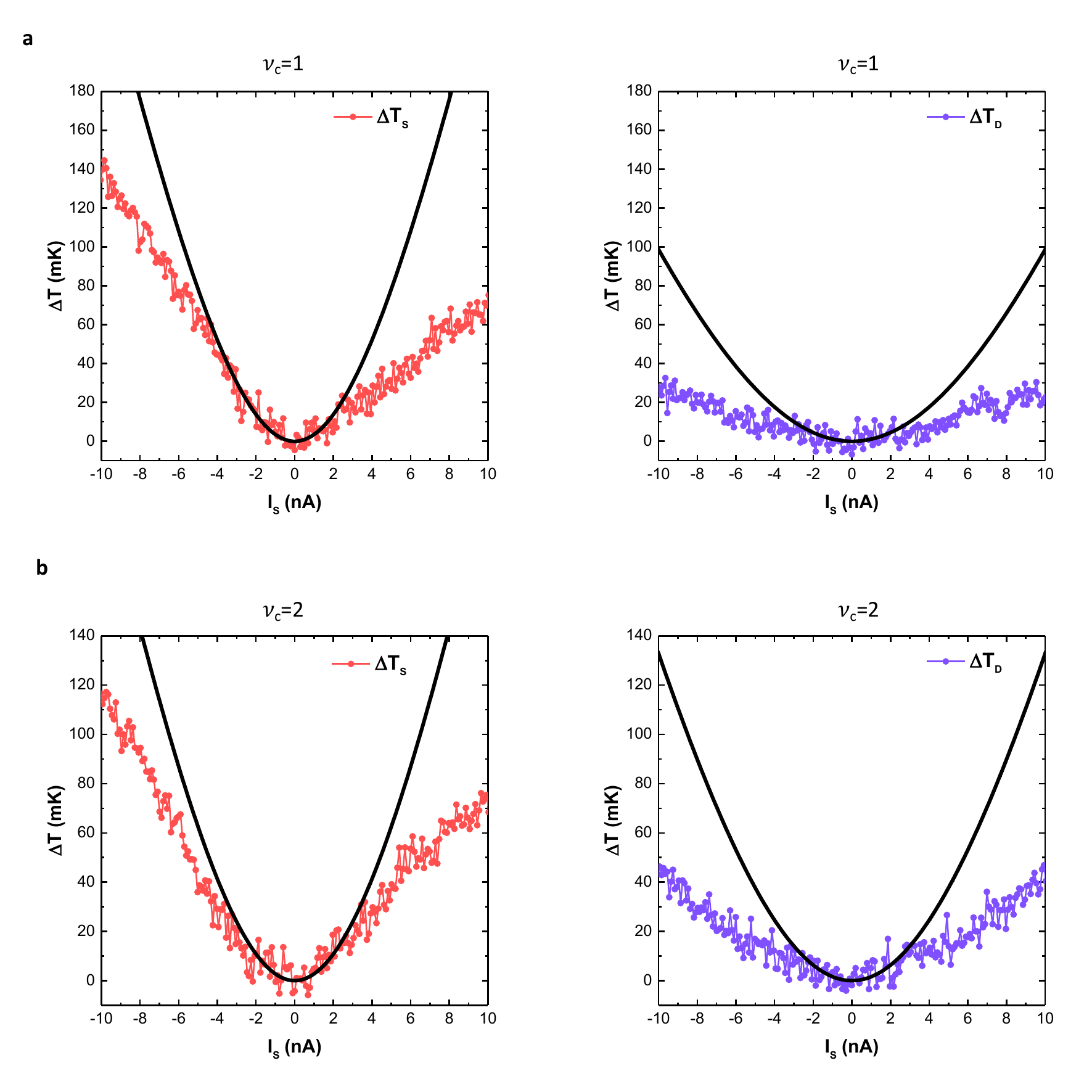}
\caption{\label{figsup-2termsample2-TvsI-Nuc=1and2} \textbf{a),} Temperature increase of the source (left, red symbols) and the detector (right, blue symbols) as a function of $\Is$ at $\nuc=1$. \textbf{b),} Temperature increase of the source (left, red symbols) and the detector (right, blue symbols) as a function of $\Is$ at $\nuc=2$. In both \textbf{a,} and \textbf{b)}, the magnetic field is 14~T and the fridge base temperature is $\sim10~$mK. The black lines are the predictions based on the calculation presented above, for $T_0\approx 310~$mK. }
\end{figure*}

The results of heat transport measurements performed at $\nuc=0$ (with $\nu=2$) for $B=14~$T are shown in Supplementary Fig.~\ref{figsup-2termsample2-TvsI10mK}. As for device 1, the detector does not show any temperature increase at $\nuc=0$ as the source temperature is increased, and vice-versa. While this qualitatively confirms our results of a vanishing bulk heat flow at $\nuc=0$, quantitatively, the temperature increase of the source (resp. detector) with $\Is$ (resp. $\Id$) is smaller than the one expected from electronic heat balance at $\nu=2$. Furthermore, the thermal rounding is much larger than expected from the base fridge temperature of about 10~mK, with the black line in Supplementary Fig.~\ref{figsup-2termsample2-TvsI10mK} corresponding to $T_0\approx310~$mK. This very large base temperature was confirmed by measurements of the equilibrium noise as a function of the back-gate voltage. We performed similar measurements at a fridge base temperature of 100~mK, with similar thermal noise amplitudes and thermal rounding as the 10~mK data, as shown in Supplementary Fig.~\ref{figsup-2termsample2-TvsI100mK}. We were not able to unambiguously identify the cause of this very large $T_0$, nor that of the marked asymmetry of the data with both $\Is$ and $\Id$. As stated above, the noise measurements for this device were performed with a different noise measurement setup, that had not been fully optimized. It is possible that the shunting resistors installed at low temperature on the dc current feed lines generate very large fluctuations (either due to the power dissipated in them by the dc and ac currents, or through poorly anchored wires vibrating in the magnetic field) that heat up the floating metal contacts. A large $T_0$ would also lead to a much larger contribution of the electron-phonon cooling in the contacts, explaining the lower thermal noise amplitude at finite $I_{\mathrm{S}/\mathrm{S}}$. Because of this, the measurements at finite $\nuc$ were not quantitatively conclusive beyond the fact that a increase in the source temperature clearly leads to an increase in the detector temperature, see Supplementary Fig.~\ref{figsup-2termsample2-TvsI-Nuc=1and2}. Nevertheless, our observation on the bulk heat flow at $\nuc=0$ are validated in this non-aligned sample.

\section{Phase diagram of the $\nu=0$ quantum Hall ferromagnet}

Our interpretation of the results is based on the framework developed in \cite{Kharitonov2012}, also used to derive the thermal conductance of the collective modes at $\nu=0$ in bilayer graphene \cite{Pientka2017}. In this framework, the four phases (CAF, F, KD, FSP) are parametrized by two anisotropies $u_\mathrm{z}$ and $u_\perp$, which describe how lattice-scale Coulomb interactions tend to polarize the ground state in a single sublattice (large negative $u_\mathrm{z}$) or in a superposition of the two sublattices (large negative $u_\perp$). The phase diagram is shown in Supplementary Fig.~\ref{figsup-phasediagram}, displaying the transition lines between each phase as black dashed line that join at a quadruple point with coordinates $(-\frac{E_\mathrm{Z}}{2},-\frac{E_\mathrm{Z}}{2})$, with $E_\mathrm{Z}$ Zeeman energy. The KD-CAF transition line has an $y=-x$ asymptotical behavior away from the quadruple point, and acquires a finite curvature as it approaches the quadruple point.

\begin{figure*}[h!]
\centering
\includegraphics[width=0.45\textwidth]{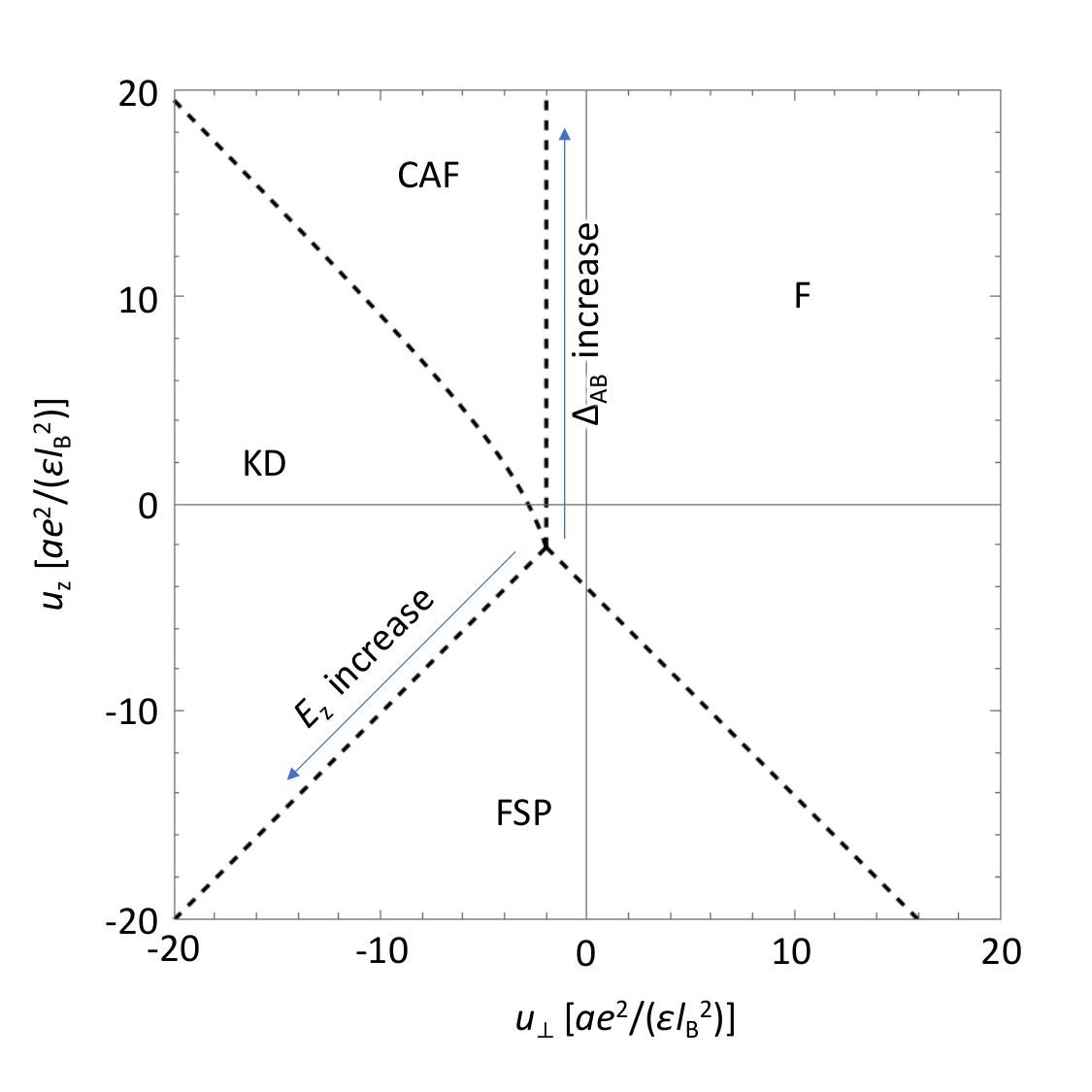}
\caption{\label{figsup-phasediagram} Phase diagram of the $\nu=0$ quantum Hall ferromagnet in monolayer graphene, adapted from Ref.~\cite{Kharitonov2012}. The dashed lines indicate the phase transition lines. The blue arrows indicate how the quadruple point shifts as the Zeeman energy $E_\mathrm{Z}$ or the zero-field sublattice gap $\Delta_\mathrm{AB}$ are increased.}
\end{figure*}

Since $u_\mathrm{z}$, $u_\perp$ and $E_\mathrm{Z}$ all scale linearly with the perpendicular magnetic field $B_\perp$, changing its value does not induce any phase transition. This makes the experimental determination of $u_\mathrm{z}$ and $u_\perp$ challenging. Several external parameters nonetheless allow inducing phase transitions. First, increasing the Zeeman energy (with either a strong in-plane magnetic field~\cite{Young2014}, or by using a high dielectric constant substrate to screen interactions~\cite{Veyrat2020}) shifts the quadruple point downwards diagonally towards negative $u_\mathrm{z}$ and $u_\perp$ along the KD-FSP transition line. At fixed $B_\perp$ and thus constant $u_\mathrm{z}$ and $u_\perp$, this induces transitions towards the F phase, as observed in refs.~\cite{Young2014,Veyrat2020}. Second, aligning the graphene flake with its hBN substrate breaks sublattice symmetry, inducing a gap $\Delta_\mathrm{AB}$ at zero magnetic field at the charge neutrality point (see above). Increasing this gap shifts the quadruple point upwards along the vertical CAF-F transition line~\cite{Zibrov2018}, in a manner similar to the effect of an electric displacement field in bilayer graphene~\cite{Kharitonov2012a}, such that the coordinates of the quadruple point are $(-\frac{E_\mathrm{Z}}{2},\Delta_\mathrm{AB}-\frac{E_\mathrm{Z}}{2})$. Since $\Delta_\mathrm{AB}$ does not depend on the magnetic field, increasing the latter has the opposite effect on the quadruple point, shifting it downwards vertically until $\Delta_\mathrm{AB}$ becomes negligible. This then induces transition from the FSP phase towards the KD and CAF phases. As explained in Ref.~\cite{Zibrov2018}, the magnetic field at which the transition occurs critically depends on the alignment angle. For samples close to alignment, this field is in the 30~T range, inaccessible in our experiments ; as the alignment angle is increased, $\Delta_\mathrm{AB}$ drops sharply, and the transition occurs at small magnetic fields, such that in practice it is not observed.

Our devices are all based on monolayer graphene flakes encapsulated in $\sim40~$nm-thick hBN crystals, with a top and bottom gate, giving them an electrostatic environment very similar to that of the samples measured in Ref.~\cite{Zibrov2018}. We can thus expect the values of the anisotropies $u_\mathrm{z}$ and $u_\perp$ to be similar: $u_{\mathrm{z}}\approx15\frac{a e^2}{\epsilon l_\mathrm{B}^2}$ and $u_\perp\approx-10\frac{a e^2}{\epsilon l_\mathrm{B}^2}$ with $a$ the interatomic distance in graphene, and $l_\mathrm{B}$ the magnetic length given by the perpendicular magnetic field. The aligned 2-terminal device (device 1) is thus expected to be in the FSP phase up to about $13-19~$T (such that the data at 14~T could be close to the FSP-KD transition, and the two other, non-aligned devices should be in the KD or CAF phase.

\section{Magnitude of the heat flow in the CAF phase for bilayer graphene}

The magnitude of the heat flow for the CAF phase was recently calculated for bilayer graphene~\cite{Pientka2017}, for typical sample dimensions (width $W=5~\mu$m) comparable to ours: $J_\mathrm{CAF}\approx6\times 10^{-12}(T_\mathrm{S}^3-T_\mathrm{0}^3)$. Because of its $T^3$ dependence, its comparison to the quantized heat flow for a single electronic channel $\Jqe=\frac{\kappa_0}{2}(T_\mathrm{S}^2-T_\mathrm{0}^2)$ strongly depends on both $\Ts$ and $T_0$. We plot in Supplementary Fig.~\ref{figsup-Jcaf_calc} the calculated ratio $J_\mathrm{CAF}/\Jqe$ as a function of $\DeltaTs$ for various values of $T_0$ corresponding to the ranges explored in our measurements. The calculation shows a thermal rounding corresponding to $T_0$, beyond which the ratio becomes linear when $\Ts\gg T_0$. Even at the lowest $T_0=10~$mK, the ratio is at its smallest equal to $0.2$, and reaches unity for $\Ts<100~$mK which is well within our measurement range.

\begin{figure*}[h]
\centering
\includegraphics[width=0.6\textwidth]{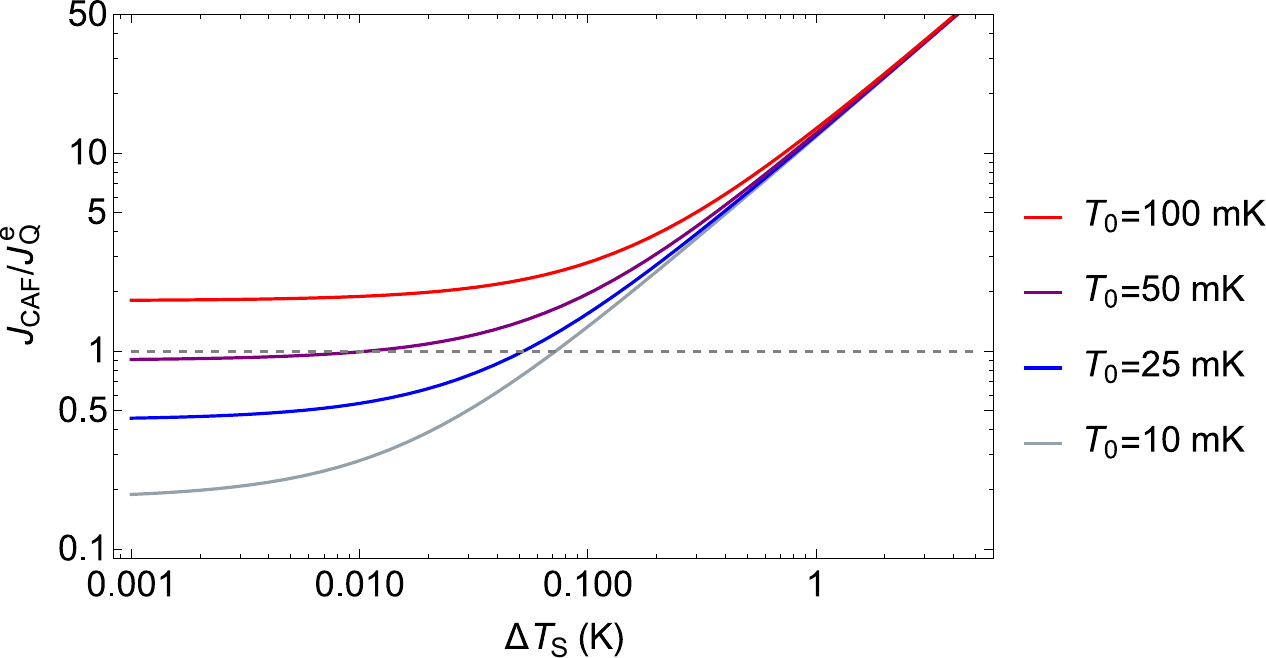}
\caption{\label{figsup-Jcaf_calc} Ratio between the predicted bulk heat flow in the CAF phase of bilayer graphene $J_\mathrm{CAF}$ and the quantized electronic heat flow for a single ballistic channel $\Jqe$, as a function of the increase in source temperature $\DeltaTs$, for $T_0$ ranging from 10 mK (blue-grey) to 100 mK (red).}
\end{figure*}

\begin{figure*}[h]
\centering
\includegraphics[width=0.85\textwidth]{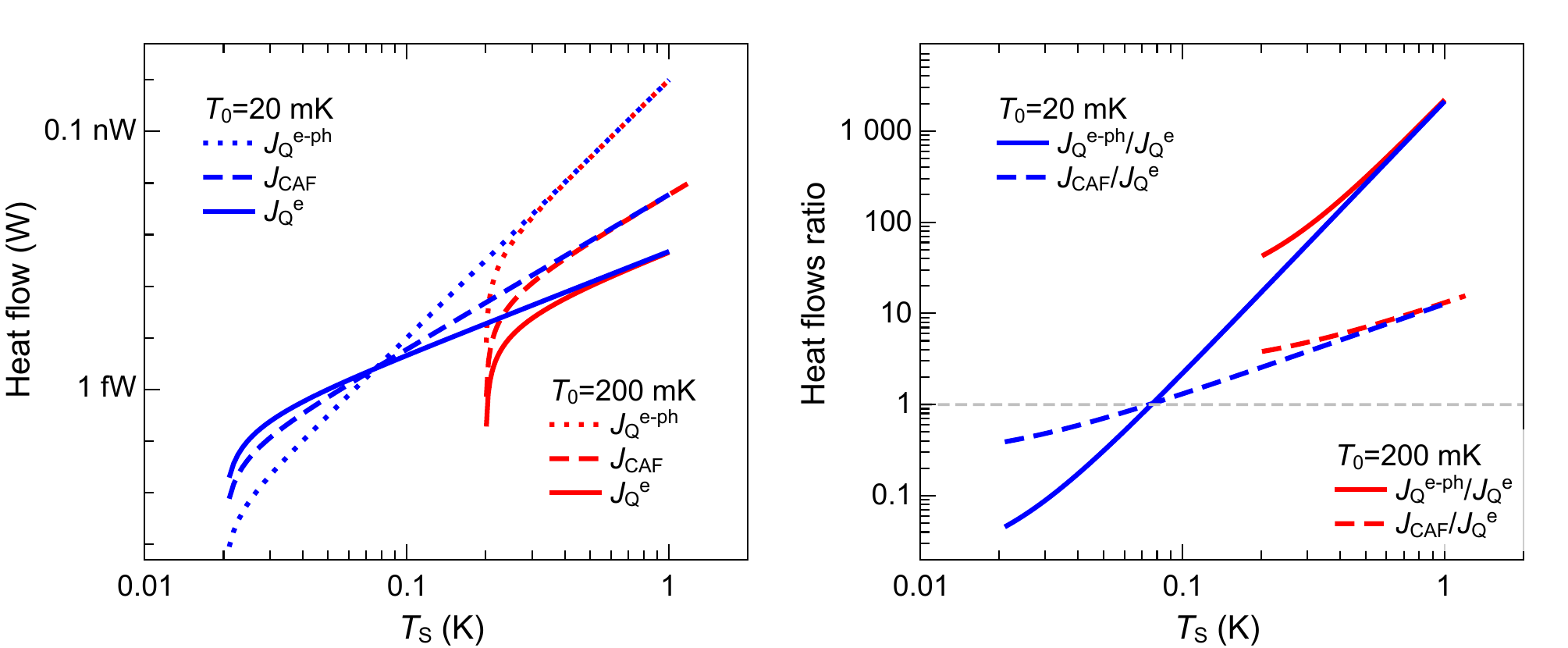}
\caption{\label{figsup-eph-e-cooling} Left: calculated heat flows carried by a single electronic channel $\Jqe$ (full line), by the gapless collective modes of the CAF phase in bilayer graphene $J_\mathrm{CAF}$ (dashed line), and due to electron-phonon coupling in the metallic island $J_{\mathrm{Q}}^\mathrm{e-ph}$ (dotted line), for $T_0=20~$mK (blue) and 200~mK (red). Right: corresponding ratios $J_{\mathrm{Q}}^\mathrm{e-ph}/\Jqe$ (full line) and $J_\mathrm{CAF}/\Jqe$ (dashed line).}
\end{figure*}

Supplementary Figure~\ref{figsup-eph-e-cooling} shows direct comparisons in log-log scale between the heat flow of a single ballistic electron channels $\Jqe$, the calculated CAF phase heat flow $J_\mathrm{CAF}$, and the heat typical flow due to electron-phonon coupling $J_{\mathrm{Q}}^\mathrm{e-ph}\approx10^{-9}(T_\mathrm{S}^5-T_\mathrm{0}^5)$ \cite{Jezouin2013a}. They are calculated as a function of $\Ts$ for two ranges of $T_0$. At low $T_0=20~$mK, there is a substantial range of $\Ts$, up to about 100~mK, for which the electronic heat flow dominates, while the bulk heat flow still has comparable amplitudes. As $\Ts$ is increased, the electron-phonon cooling largely becomes dominant, overcoming the other contributions by several orders of magnitude. At large $T_0=200~$mK, $J_{\mathrm{Q}}^\mathrm{e-ph}$ is immediately dominant, such that even though $J_\mathrm{CAF}$ is about ten times larger than $\Jqe$, it is completely obscured by the electron-phonon cooling. From these approximate numbers (the amplitude of $J_{\mathrm{Q}}^\mathrm{e-ph}$ is about two to three times smaller in our samples~\cite{LeBreton2022}, since the floating metallic contacts generally have smaller areas and thicknesses than the Au/Ge/Ni ohmic contacts used in GaAs experiments~\cite{Jezouin2013a}), we can conclude that it is better to keep $T_0$ as low as possible and to explore a wide range of $\Ts$.

\section{Spin/valley waves confinement and spectrum quantization}

A recent experiment in bilayer graphene~\cite{Fu2021} reports the observation of non-local interference patterns in the electrical conductance across a region of the sample set to $\nu=0$. These patterns are interpreted as the signature of Fabry-Perot interferences of the gapless spin waves of the $\nu=0$ CAF state in bilayer graphene. This raises the question whether the absence of heat flow at $\nuc=0$ in our experiments (particularly in the heat Corbino device) could stem from mode quantization in the cavities formed by the sample. Indeed, in the Fabry-Perot geometry  of Ref.~\cite{Fu2021}, where the $\nu=0$ region is a narrow strip enclosed between two $\nu=4$ regions (the distance between the two $\nu=0 / \nu=4$ interface is about 2 micrometers, and the width of both interfaces is about 4 micrometers), the mode quantization gaps are typically $60~\mu$eV, much higher than the explored temperature range in our experiment (except for the data shown in main text Fig.~4). We can estimate the magnitude of the gaps for the CAF phase in monolayer graphene using $E=\hbar v_\mathrm{CAF} \pi/L$, with $v_\mathrm{CAF}$ and $L$ the CAF spin wave velocity, and the typical dimension of the cavity formed by the sample, respectively. With $v_\mathrm{CAF}\approx 226~$km$/$s (extracted from Ref.~\cite{Wei2021}) this yields a large gap of about $5.4~$K, larger than those in bilayer, due to the higher spin waves velocity. While this could explain our results, we point out the fact that no signature of spin/valley wave interference at $\nu=0$ has been reported in monolayer so far~\cite{Wei2018,Stepanov2018}. This might be due to the recently observed domains of Kekule bond order~\cite{Coissard2022} yielding diffusive-like valley wave transport in the expected KD phase. Similarly, the coexistence between CAF and KD phases discussed in Ref.~\cite{Das2022} is likely to hinder mode quantization in monolayer graphene.

\section{Thermal interface resistance}

The bulk heat flow predicted for the CAF phase in Ref.~\cite{Pientka2017} was calculated using a principle very close to that of our experiment: the hot and cold sources are electron baths located in metallic leads connected to the (bilayer) graphene flake. The baths impose their respective temperatures to the collective modes on either side of the sample: in other words, the authors do not consider any effective thermal interface resistance between the electron in the metallic leads and the bulk collective modes. 
The existence of this interface resistance however cannot be excluded from our experiments; nonetheless, we may expect a better coupling of the electron bath to the collective modes than to phonons in the metallic island. This can be intuitively understood by the fact that the collective modes are essentially electron-hole type excitations. In a recent work~\cite{Assouline2021}, we showed that the magnonic collective excitations at $\nu=1$ have a finite dipole due to their electron-hole nature, allowing them to couple to other electronic degrees of freedom. Furthermore, the metallic contacts used in our experiments tend to locally n-dope the graphene flake, a fact highlighted by the ability to generate and detect magnons at $\nuc=1$, as explained in Ref.~\cite{Wei2018}. Thus, the reservoir of hot electrons actually extends into the graphene flake itself, potentially increasing the coupling to collective modes.

\bibliography{nu0}